\documentclass[journal,10pt,letterpaper,final,twocolumn]{IEEEtran}%
\usepackage{amsmath}
\usepackage{amsfonts}
\usepackage{cite}
\usepackage{amssymb}
\usepackage{mathrsfs}
\usepackage{graphicx}
\usepackage[usenames,dvipsnames]{color}
\usepackage{epstopdf}
\usepackage[all]{xy}
\usepackage[]{mcode}
\usepackage{supertabular}
\usepackage{subcaption}
\usepackage{array}
\usepackage{diagbox}

\providecommand{\U}[1]{\protect\rule{.1in}{.1in}}
\pdfoutput=1
\providecommand{\U}[1]{\protect\rule{.1in}{.1in}}

\newcolumntype{L}[1]{>{\raggedright\let\newline\\\arraybackslash\hspace{0pt}}m{#1}}
\newcolumntype{C}[1]{>{\centering\let\newline\\\arraybackslash\hspace{0pt}}m{#1}}
\newcolumntype{R}[1]{>{\raggedleft\let\newline\\\arraybackslash\hspace{0pt}}m{#1}}

\newcommand{\qed}{\nobreak \ifvmode \relax \else
      \ifdim\lastskip<1.5em \hskip-\lastskip
      \hskip1.5em plus0em minus0.5em \fi \nobreak
      \vrule height0.75em width0.5em depth0.25em\fi}

\begin{document}

\title{Enhanced Orthogonal Frequency-Division Multiplexing with Subcarrier Number Modulation}
\author{Shuping Dang, \textit{Member, IEEE}, Guoqing Ma, \textit{Student Member, IEEE}, Basem Shihada, \textit{Senior Member, IEEE}, Mohamed-Slim Alouini, \textit{Fellow, IEEE}
  \thanks{S. Dang, G. Ma, B. Shihada, and M.-S. Alouini are with Computer, Electrical and Mathematical Science and Engineering Division, King Abdullah University of Science and Technology (KAUST), 
Thuwal 23955-6900, Kingdom of Saudi Arabia (e-mail: \{shuping.dang, guoqing.ma, basem.shihada, slim.alouini\}@kaust.edu.sa).

}}

\maketitle

\begin{abstract}
A novel modulation scheme termed orthogonal frequency-division multiplexing with subcarrier number modulation (OFDM-SNM) has been proposed and regarded as one of the promising candidate modulation schemes for next generation networks. Although OFDM-SNM is capable of having a higher spectral efficiency (SE) than OFDM with index modulation (OFDM-IM) and plain OFDM under certain conditions, its reliability is relatively inferior to these existing schemes, because the number of active subcarriers varies. In this regard, we propose an enhanced OFDM-SNM scheme in this paper, which utilizes the flexibility of placing subcarriers to harvest a coding gain in the high signal-to-noise ratio (SNR) region. In particular, we stipulate a methodology that optimizes the subcarrier activation pattern (SAP) by subcarrier assignment using instantaneous channel state information (CSI) and therefore the subcarriers with higher channel power gains will be granted the priority to be activated, given the number of subcarriers is fixed. We also analyze the proposed enhanced OFDM-SNM system in terms of outage and error performance. The average outage probability and block error rate (BLER) are derived and approximated in closed-form expressions, which are further verified by numerical results generated by Monte Carlo simulations.  The high-reliability nature of the enhanced OFDM-SNM makes it a promising candidate for implementing in the Internet of Things (IoT) with stationary machine-type devices (MTDs), which are subject to slow fading and supported by proper power supply.
\end{abstract}


\section{Introduction}
\IEEEPARstart{B}{ecause} of the saturation of base station (BS) deployments in fourth generation (4G) networks, it becomes increasingly difficult to enhance the spectral efficiency (SE) of wireless communication by spatial optimization and further densifying networks \cite{8370884}. To cope with the increasingly high demand for data throughput, many researchers resort to novel modulation schemes. In this regard, a variety of novel modulation schemes were proposed. In the space domain, for multiple-input and multiple-output (MIMO) systems, spatial modulation (SM) and space-shift keying (SSK) were introduced to utilize the indices of transmit antennas to convey additional information bits \cite{4382913,5165332}. Although helpful, SM and SSK supported by a multi-antenna architecture will inevitably result in higher system complexity and larger device size. However, with the advancement of the Internet of Things (IoT) and machine-type communication (MTC) networks, communication nodes are miniaturized and simple, which might not be able to afford such a high-complexity structure yielded by SM and SSK \cite{6678765}. 

Subcarrier-index modulation (SIM) orthogonal frequency-division multiplexing (OFDM) was proposed as the first attempt to extend the gist of SM to the frequency domain in order to solve the aforementioned issues regarding system complexity and device size. There are two different versions of SIM OFDM proposed in \cite{5449882} and \cite{6162549}, respectively. However, the former relies on a cross-layer design based on forward error control techniques, and the latter has a lower transmission rate, which are impractical for general cases. The first widely recognized success to transplant the gist of SM to the frequency domain refers to the OFDM with index modulation (OFDM-IM) \cite{6587554}. By OFDM-IM, a new dimension of subcarrier index is employed for modulating extra bits in addition to classic phase and amplitude dimensions of the signal constellation. The proper feasibility and high efficiency of OFDM-IM have then drawn the attention from industry and academia and sparked the research enthusiasm since 2013 until now\footnote{From a taxonomic viewpoint, after the concept of OFDM-IM gets well-known, SIM OFDM and OFDM-IM are sometimes regarded as synonyms and used exchangeably  \cite{7469311,8358694}.} \cite{7509396,8004416,8353362,8417419,8315127}. Despite the feasibility in practical scenarios, OFDM-IM has several drawbacks. First, by OFDM-IM, the number of active subcarriers in each transmission attempt is fixed to a given number, which restricts the improvement of the SE of OFDM-IM. Meanwhile, the codebook design of OFDM-IM depending on either a look-up table or the combinatorial method is of high complexity and has not fully exploited the frequency selectivity for reliability enhancement \cite{8101465}.

In order to cope with the aforementioned drawbacks of OFDM-IM, a novel modulation scheme termed OFDM with subcarrier number modulation (OFDM-SNM) was proposed and preliminarily investigated in terms of SE, error performance and energy efficiency (EE) in \cite{8362748}.  In essence, OFDM-SNM can be regarded as a `sibling' modulation scheme sharing a similar nature with OFDM-IM, which relies on another set of subcarrier activation patterns (SAPs) and a unique information mapping relation. Technically different from OFDM-IM, by OFDM-SNM, the numbers of active subcarriers in each transmission round are utilized to convey extra bits, instead of the indices of active subcarriers. In this way, a new \textit{active subcarrier number} (ASN) dimension can be employed to convey additional information. Primary results illustrated in \cite{8362748} have shown that a higher SE is achievable by OFDM-SNM than those of OFDM-IM and plain OFDM when binary phase-shift keying (BPSK) is in use for amplitude phase modulation (APM) on individual subcarriers. Also, EE and reliability measured by error performance are shown to be better than those of plain OFDM and comparable to those yielded by OFDM-IM. Although verified by neither analytical nor numerical results, a hypothesis is given in \cite{8362748} that there is a potential to enhance the system reliability of OFDM-SNM by the flexibility of placing active subcarriers because of the frequency selectivity. This results in an opportunity to incorporate some channel-dependent adaptation mechanisms in OFDM-SNM to further enhance the system reliability, just as for other multi-carrier system paradigms \cite{4769396,kocan2010performance,7462218,8093591,8241721,furqan2018adaptive,8519769}.

In this regard, we propose an enhanced OFDM-SNM scheme in this paper, which is supported by subcarrier assignment. In particular, we consider a slow fading environment and the subcarriers with better quality, i.e., higher instantaneous channel power gains will be granted the priority for use by the proposed enhanced OFDM-SNM scheme. Therefore, with the help of instantaneous channel state information (CSI), an adaptive modulation mechanism is formed, which can provide a dynamic codebook and enhance the performance of OFDM-SNM by a coding gain. Apart from this all-important contribution, we also provide a series of in-depth performance analysis and comparisons with original OFDM-SNM, aiming at supplementing the primary results given in \cite{8362748}. Specifically, we determine the transmission rate of OFDM-SNM in bit per channel use (bpcu) and investigate the outage and error performance of enhanced OFDM-SNM by average outage probability and average block error rate (BLER), respectively. All analytical results are derived or approximated in closed form and verified by numerical results generated by Monte Carlo simulations. The high-reliability nature of enhanced OFDM-SNM particularly suits the applications in the IoT with stationary machine-type devices (MTDs), which are subject to slow fading and supported by proper power supply.

The rest of this paper is organized as follows. The system model of enhanced OFDM-SNM is detailed in Section \ref{sm}, in which we also present some relevant information regarding transmission rate. Then, the outage and error performance are analyzed in Section \ref{opa} and Section \ref{epa}, respectively. To support the analytical derivations and provide performance comparisons with the original OFDM-SNM, numerical results are presented and discussed in Section \ref{nr}. Finally, we conclude the paper in Section \ref{c}. Readers who are interested in the transmission rate comparison among OFDM-SNM, OFDM-IM, and plain OFDM would also find Appendix useful. Also, for readers' convenience, we list the key notations and abbreviations in Table \ref{listnotation} and Table \ref{listabbrev}, respectively.

\begin{table}[!t]
\renewcommand{\arraystretch}{1.3}
\caption{List of key notations used in this paper.}
\label{listnotation}
\centering
\begin{tabular}{|C{1cm}|L{6cm}|}
\hline
Notation & Definition/explanation\\
\hline
$h(n)$ & Channel coefficient of the $n$th subcarrier\\
$k$ & Index of SAP\\
$M$ & Amplitude phase modulation order\\
$N$ & Number of subcarriers\\
$N_0$ & Average noise power\\
$n$ & Index of subcarrier\\
$\overline{P}_e$ & Average block error rate\\
$P_t$ & Total transmit power\\
$p_H$ & Length of heading bit stream\\
$p_S(k)$ & Length of subsequent bit stream of the $k$th SAP\\
$p(k)$ & Length of entire bit stream of the $k$th SAP\\
$p_{\mathrm{IM}}$ & Transmission rate of OFDM-IM\\
$p_{\mathrm{OFDM}}$ & Transmission rate of plain OFDM\\
$\bar{p}$ & Average transmission rate in bpcu\\
$T$ & Number of active subcarriers predefined by OFDM-IM\\
$T(k)$ & Number of active subcarriers of the $k$th SAP\\
$\chi_n$ & Complex constellation symbol conveyed on the $n$th active subcarrier\\
$\mu$ & Average channel power gain\\
$\overline{\Phi}$ & Average outage probability\\
$\xi$ & Preset outage threshold\\
\hline
\end{tabular}
\end{table}

\begin{table}[!t]
\renewcommand{\arraystretch}{1.3}
\caption{List of abbreviations used in this paper.}
\label{listabbrev}
\centering
\begin{tabular}{|C{1cm}|L{6cm}|}
\hline
Abbr. & Definition/explanation\\
\hline
APM & Amplitude phase modulation\\
ASN & Active subcarrier number\\
AWGN & Additive white Gaussian noise\\
BER & Bit error rate\\
BLER & Block error rate\\
bpcn & Bit per channel use\\
BPSK & Binary phase-shift keying\\
BS & Base station\\
CDF & Cumulative distribution function\\
CP & Cyclic prefix\\
CR & Cognitive radio\\
CSI & Channel state information\\
CSM & Channel state matrix\\
EE & Energy efficiency\\
IFFT & Inverse fast Fourier transform\\
i.i.d. & Independent and identically distributed\\
IM & Index modulation\\
IoT & Internet of Things\\
LTE & Long-Term Evolution\\
MIMO & Multiple-input and multiple-output\\
ML & Maximum-likelihood (detection)\\
MTC & Machine-type communication\\
MTD & Machine-type device\\
OFDM & Orthogonal frequency-division multiplexing\\
PDF & Probability density function\\
PEP & Pairwise error probability\\
PSK & Phase-shift keying\\
QAM & Quadrature amplitude modulation\\
SAP & Subcarrier activation pattern\\
SE & Spectral efficiency\\
SIM & Subcarrier-index modulation\\
SM & Spatial modulation\\
SNM & Subcarrier number modulation\\
SNR & Signal-to-noise ratio\\
SSK & Space-shift keying\\
4G & Fourth generation (networks)\\
\hline
\end{tabular}
\end{table}

\section{System Model}\label{sm}
\subsection{System Framework}
In this paper, we consider a simplistic point-to-point multi-carrier communication scenario supported by OFDM architecture, and focus on only one single group of $N$ subcarriers without loss of generality. In modern multi-carrier systems, these $N$ subcarriers are generated by taking the fast inverse fast Fourier transform (IFFT) with  insertion of sufficiently long cyclic prefix (CP) and can thereby operate mutually independently without interference and correlation \cite{1273689}. We denote the set of subcarriers as $\mathcal{N}$. By involving OFDM-SNM, the functionality of subcarrier is not only to convey data constellation symbols, but also to provide a unique SAP to transmit extra bits. Specifically, a subset of subcarriers $\mathcal{N}(k)$ are selected from the full set $\mathcal{N}$ to be activated, where $k$ denotes the index of a unique SAP, and the cardinality $T(k)=|\mathcal{N}(k)|$, i.e., the number of active subcarriers is utilized to modulate the heading bit sequence with a fixed length $p_H$. The relation between $p_H$ and $N$ can be easily determined by $p_H=\lfloor\log_2(N)\rfloor$, where $\lfloor\cdot\rfloor$ is the floor function and can be removed if and only if $N$ is a power of two. Having determined $\mathcal{N}(k)$, we resort to the conventional $M$-ary phase-shift keying ($M$-PSK) to convey data constellation symbols on active subcarriers\footnote{The reason for employing $M$-PSK instead of $M$-ary quadrature amplitude modulation ($M$-QAM) in this paper is because of its constant-envelope property and rotational symmetry \cite{6142142,7330022}.}, where $M$ is the APM order. These data constellation symbols are determined by a $k$-dependent \textit{subsequent} bit sequence with a variable length $p_S(k)=T(k)\log_2(M)$. As a result of the variable-length subsequent bit sequence, the entire bit sequence for modulation also has a variable length, which is $p(k)=p_H+p_S(k)$. We can average $p(k)$ over all SAPs and determine the average transmission rate in bpcu by
\begin{equation}\label{sdkasjkdj2}
\begin{split}
\bar{p}=p_H+\underset{k}{\mathbb{E}}\left\lbrace p_S(k)\right\rbrace=\lfloor\log_2(N)\rfloor+\frac{1+2^{\lfloor\log_2(N)\rfloor}}{2}\log_2(M),
\end{split}
\end{equation}
where $\mathbb{E}\{\cdot\}$ is the expected value of the enclosed random variable. For simplicity, (\ref{sdkasjkdj2}) can be reduced to 
\begin{equation}\label{sdkasjkdj10000}
\begin{split}
\bar{p}=\log_2(N)+\frac{N+1}{2}\log_2(M),
\end{split}
\end{equation}
when $N$ is a power of two (a common assumption for modern multi-carrier systems \cite{1177182}). The average transmission rate in bpcu is a key measurement for the SE of both coded and uncoded OFDM-SNM systems. As an elaborate discussion regarding the average transmission rate is lacking in \cite{8362748}, we provide a comprehensive comparison among the data transmission rates of OFDM-SNM, OFDM-IM, and plain OFDM in Appendix Note that, although the length of subsequent bit sequence $p_S(k)$ is associated with the heading bit sequence, we assume all bits are equiprobable and uncorrelated for information-theoretically maximizing the system usage.

\subsection{Signal Transmission and CSI-Based Coding}

In order to express a SAP, we employ the $k$-dependent activation state vector expressed as $\mathbf{S}(k)=[s(k,1),s(k,2),\dots,s(k,N)]^T\in\{0,1\}^{N\times 1}$, where $(\cdot)^T$ represents the matrix/vector transpose and $s(k,n)=\begin{cases}
1,~~~~\mathrm{if~the~}n\mathrm{th~subcarrier~is~active}\\
0,~~~~\mathrm{if~the~}n\mathrm{th~subcarrier~is~inactive}\\
\end{cases}$. Different from original OFDM-SNM proposed in \cite{8362748}, by which $\mathbf{S}(k)$ is completely dependent on the $p_H$-bit heading sequence, $\mathbf{S}(k)$ by the proposed enhanced OFDM-SNM is dependent on both of the $p_H$-bit heading sequence and instantaneous CSI when $T(k)< N$. Specifically, because indices of active subcarriers do not matter in OFDM-SNM, whereas number does, this provides a flexibility to activate subcarriers according to their channel qualities for a given SAP $k$, as long as the total number of active subcarriers is given. In this regard, subcarrier assignment can be involved to select appropriate subcarriers to activate based on instantaneous CSI, so as to generated a coded mapping scheme from incoming bit sequences to SAPs and attain a coding gain. In particular, when $T(k)< N$, we stipulate the rule to generate subset $\mathcal{N}(k)$ and assign $T(k)$ active subcarriers by the criterion below\footnote{This subcarrier assignment criterion is equivalent to selecting the $T(k)$ subcarriers from all $N$ subcarriers with the first to the $T(k)$th largest instantaneous channel power gains.}:
\begin{equation}\label{dasjd22226a}
\begin{split}
\mathcal{N}(k)=\underset{\begin{subarray}{c}\tau\subset\mathcal{N},~|\tau|=T(k)\end{subarray}}{\arg\max}\left\lbrace \sum_{n\in\tau}|h(n)|^2\right\rbrace,
\end{split}
\end{equation}
where $h(n)$ is the complex channel coefficient of the $n$th subcarrier and $|h(n)|^2$ is thereby the corresponding channel power gain; $\tau$ is an arbitrary subset of active subcarriers that has a cardinality of $T(k)$.

\begin{figure}[!t]
\centering
\includegraphics[width=3.5in]{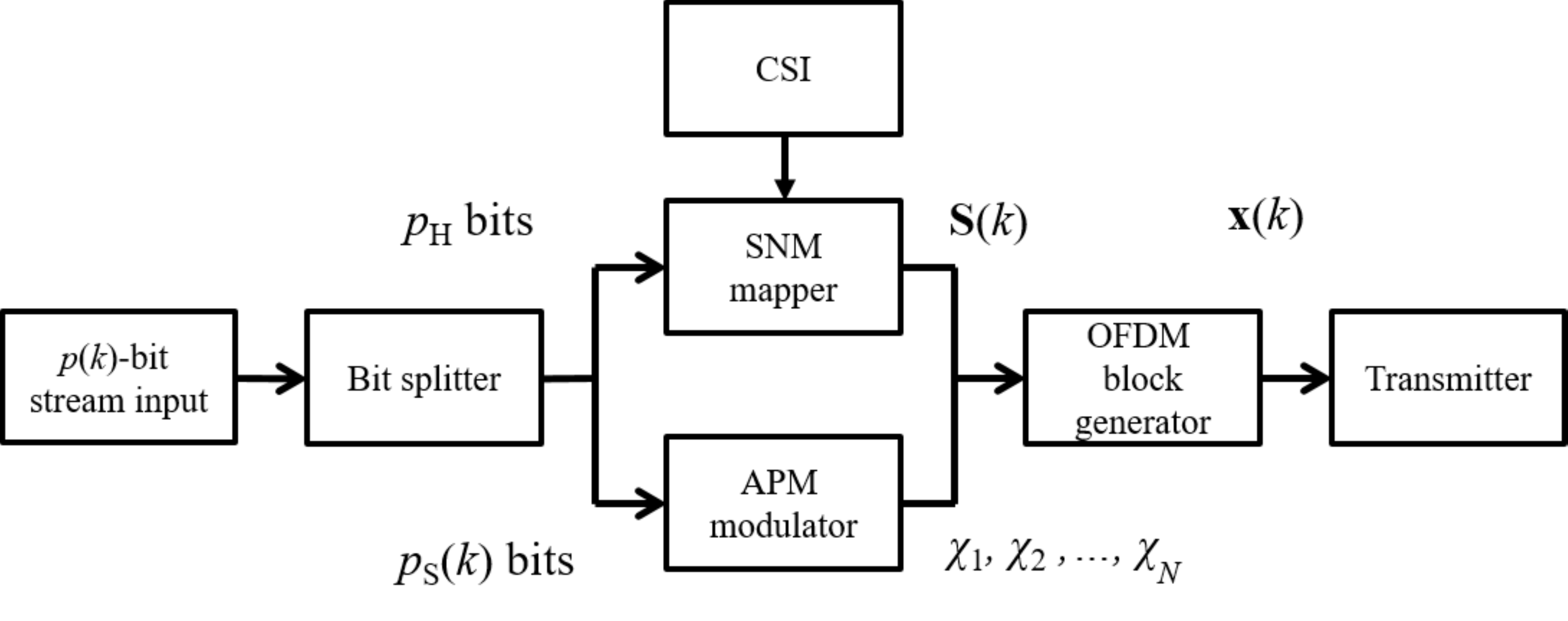}
\caption{Enhanced OFDM-SNM transmitter framework (for a single OFDM block).}
\label{sys}
\end{figure}

Then, with the optimized $\mathbf{S}(k)$ and $p_S(k)$-bit subsequent sequence, IFFT can be employed to generate the OFDM block for transmission just as plain OFDM, which gives $\mathbf{x}(k)=[x(k,1),x(k,2),\dots,x(k,N)]^T\in\mathbb{C}^{N\times 1}$, where $x(k,n)=\begin{cases}
\chi_n,~~~~\mathrm{if}~n\in\mathcal{N}(k)\\
0,~~~~~~\mathrm{otherwise}
\end{cases}$ and $\chi_n$ is the complex constellation symbol conveyed on the $n$th active subcarrier. Without loss of generality, we normalize it by $\chi_n\chi_n^*=1$ for simplicity. A complete framework of the enhanced OFDM-SNM transmitter is illustrated in Fig. \ref{sys} for clarity. To illustrate the modulation and coding procedures, we give an example with $N=4$ (with four subcarriers in total for a single subcarrier group) and $M=2$ (BPSK is in use) infra.

\subsubsection*{An example}
Given the instantaneous channel power gains $\{|h(1)|^2,|h(2)|^2,|h(3)|^2,|h(4)|^2\}=\{1.6583,0.3361,3.1437,0.8722\}$, it is straightforward to have $|h(3)|^2>|h(1)|^2>|h(4)|^2>|h(2)|^2$, which yields the priority among four subcarriers. Consequently, for $T(k)=1$, we should activate subcarrier 3 due to its largest channel power gain. For $T(k)=2$, because $|h(1)|^2+|h(3)|^2$ is the largest sum compared to other five combinations, we should activate subcarriers 1 and 3. Similarly for $T(k)=3$, because $|h(1)|^2+|h(3)|^2+|h(4)|^2$ is the largest sum compared to other three combinations, we should activate subcarriers 1, 3 and 4. Finally, when $T(k)=N=4$, as all subcarriers are required to be activated, no subcarrier assignment is needed anymore. Therefore, we finally have the optimized/coded mapping relation between incoming bit sequences and SAPs in Table \ref{exmtable}.

\begin{table}[!t]
\renewcommand{\arraystretch}{1.3}
\caption{An example of the optimized/coded mapping relation table of enhanced OFDM-SNM when $N=4$ and $M=2$, given $|h(3)|^2>|h(1)|^2>|h(4)|^2>|h(2)|^2$.}
\label{exmtable}
\centering
\begin{tabular}{c|c|c|c|c|c}
\hline
$k$ & $p(k)$ & $p_H$ bits& $p_S(k)$ bits& $\mathbf{S}(k)$ & $\mathbf{x}(k)$\\
\hline\hline
1 & $3$ & $00$ & $0$ & $[0,0,1,0]^T$ & $[0,0,-1,0]^T$\\
2 & $3$ & $00$ & $1$ & $[0,0,1,0]^T$ & $[0,0,+1,0]^T$\\
3 & $4$ & $01$ & $00$ & $[1,0,1,0]^T$ & $[-1,0,-1,0]^T$\\
4 & $4$ & $01$ & $01$ & $[1,0,1,0]^T$ & $[-1,0,+1,0]^T$\\
5 & $4$ & $01$ & $10$ & $[1,0,1,0]^T$ & $[+1,0,-1,0]^T$\\
6 & $4$ & $01$ & $11$ & $[1,0,1,0]^T$ & $[+1,0,+1,0]^T$\\
7 & $5$ & $10$ & $000$ & $[1,0,1,1]^T$ & $[-1,0,-1,-1]^T$\\
8 & $5$ & $10$ & $001$ & $[1,0,1,1]^T$ & $[-1,0,-1,+1]^T$\\
9 & $5$ & $10$ & $010$ & $[1,0,1,1]^T$ & $[-1,0,+1,-1]^T$\\
10 & $5$ & $10$ & $011$ & $[1,0,1,1]^T$ & $[-1,0,+1,+1]^T$\\
11 & $5$ & $10$ & $100$ & $[1,0,1,1]^T$ & $[+1,0,-1,-1]^T$\\
12 & $5$ & $10$ & $101$ & $[1,0,1,1]^T$ & $[+1,0,-1,+1]^T$\\
13 & $5$ & $10$ & $110$ & $[1,0,1,1]^T$ & $[+1,0,+1,-1]^T$\\
14 & $5$ & $10$ & $111$ & $[1,0,1,1]^T$ & $[+1,0,+1,+1]^T$\\
15 & $6$ & $11$ & $0000$ & $[1,1,1,1]^T$ & $[-1,-1,-1,-1]^T$\\
16 & $6$ & $11$ & $0001$ & $[1,1,1,1]^T$ & $[-1,-1,-1,+1]^T$\\
17 & $6$ & $11$ & $0010$ & $[1,1,1,1]^T$ & $[-1,-1,+1,-1]^T$\\
18 & $6$ & $11$ & $0011$ & $[1,1,1,1]^T$ & $[-1,-1,+1,+1]^T$\\
19 & $6$ & $11$ & $0100$ & $[1,1,1,1]^T$ & $[-1,+1,-1,-1]^T$\\
20 & $6$ & $11$ & $0101$ & $[1,1,1,1]^T$ & $[-1,+1,-1,+1]^T$\\
21 & $6$ & $11$ & $0110$ & $[1,1,1,1]^T$ & $[-1,+1,+1,-1]^T$\\
22 & $6$ & $11$ & $0111$ & $[1,1,1,1]^T$ & $[-1,+1,+1,+1]^T$\\
23 & $6$ & $11$ & $1000$ & $[1,1,1,1]^T$ & $[+1,-1,-1,-1]^T$\\
24 & $6$ & $11$ & $1001$ & $[1,1,1,1]^T$ & $[+1,-1,-1,+1]^T$\\
25 & $6$ & $11$ & $1010$ & $[1,1,1,1]^T$ & $[+1,-1,+1,-1]^T$\\
26 & $6$ & $11$ & $1011$ & $[1,1,1,1]^T$ & $[+1,-1,+1,+1]^T$\\
27 & $6$ & $11$ & $1100$ & $[1,1,1,1]^T$ & $[+1,+1,-1,-1]^T$\\
28 & $6$ & $11$ & $1101$ & $[1,1,1,1]^T$ & $[+1,+1,-1,+1]^T$\\
29 & $6$ & $11$ & $1110$ & $[1,1,1,1]^T$ & $[+1,+1,+1,-1]^T$\\
30 & $6$ & $11$ & $1111$ & $[1,1,1,1]^T$ & $[+1,+1,+1,-1]^T$\\
\hline
\end{tabular}
\end{table}

\subsection{Signal Reception and Detection}
Propagating over parallel fading channels, the received OFDM block at the OFDM-SNM receiver can be written as
\begin{equation}
\mathbf{y}(k)=\sqrt{\frac{P_t}{T(k)}}\mathbf{H}\mathbf{x}(k)+\mathbf{w}\in\mathbb{C}^{N\times 1},
\end{equation}
where $P_t$ is the total transmit power at the OFDM-SNM transmitter, which is uniformly distributed over $T(k)$ active subcarriers; $\mathbf{H}=\mathrm{diag}\{h(1),h(2),\dots,h(N)\}$ represents the channel state matrix (CSM); $\mathbf{w}=[w(1),w(2),\dots,w(N)]^T$ is the vector of additive white Gaussian noise (AWGN) at the receiver, and $w(n)\sim\mathcal{CN}(0,N_0)$ is the AWGN sample on the $n$th subcarrier with the average noise power $N_0$.

To provide the optimal detection, we employ the maximum-likelihood (ML) detection scheme at the receiver with the detection criterion infra to decode the received OFDM block:
\begin{equation}\label{dsajhj2a223232}
\begin{split}
\hat{\mathbf{x}}(\hat{k})=\underset{\dot{\mathbf{x}}(\dot{k})\in\mathcal{X}}{\arg\min}\begin{Vmatrix}{\mathbf{y}}({k})-\sqrt{\frac{P_t}{T(\dot{k})}}\mathbf{H}\dot{\mathbf{x}}(\dot{k})\end{Vmatrix}_F,
\end{split}
\end{equation}
where $\begin{Vmatrix}\cdot\end{Vmatrix}_F$ denotes the Frobenius norm of the enclosed matrix/vector; $\mathcal{X}$ is the full set of legitimate OFDM blocks by enhanced OFDM-SNM and its cardinality is $|\mathcal{X}|=\sum_{n=1}^{N}M^n=\frac{M(M^N-1)}{M-1}$, which is also the size of search space for OFDM block detection and characterizes the detection complexity. Meanwhile, one should note that for implementing OFDM-SNM with ML detection in practice, subcarrier interleaved grouping is indispensable, which restricts the number of subcarriers $N$ for each group to a relatively small value \cite{6587554,6841601,7330022,7086323}.

Besides, owing to the normalization of the transmitted constellation symbol $\chi_n\chi_n^*=1$, the received signal-to-noise ratio (SNR) on each subcarrier is given by 
\begin{equation}\label{ueowjkawq2993}
\gamma(k,n)=\begin{cases}
\frac{P_t|h(n)|^2}{T(k)N_0},~~~~n\in\mathcal{N}(k)\\
0,~~~~~~~~~~~~~\mathrm{otherwise}
\end{cases}
\end{equation}
which is an important indicator of the receiving quality of a single active subcarrier, and can also reflect the holistic reliability of the enhanced OFDM-SNM system.

\subsection{Channel Model}
In this paper, a slow Rayleigh fading channel is assumed with the probability density function (PDF) and cumulative distribution function (CDF) with respect to the instantaneous channel power gain $|h(n)|^2$ as follows:
\begin{equation}\label{dkasjdk2pdfcdf}
f_g(\nu)=\frac{1}{\mu}\mathrm{exp}\left(-\frac{\nu}{\mu}\right)~\Leftrightarrow~F_g(\nu)=1-\mathrm{exp}\left(-\frac{\nu}{\mu}\right)
\end{equation} 
where $\mu$ is the average channel power gain that is the same for all subcarriers, which refers to the independent and identically distributed (i.i.d.) parallel fading model for multi-carrier systems\footnote{The i.i.d. parallel fading model is validated by the implementation of CP with sufficient length, perfect synchronization in both time and frequency domain as well as proper subcarrier grouping \cite{6841601}. As a consequence, a frequency-selective channel for OFDM-SNM systems can be modeled by a number of frequency-flat Rayleigh fading channels with independent channel gains \cite{6587554}. This can be justified by the block fading model in frequency akin to systems that employ a resource block frame/packet structure (e.g., LTE), and hence the assumption of independent fading in frequency holds \cite{7445895}.}. 

Besides, we also assume that fading channels comply with the slow fading model. To be specific, the slow or quasi-static attribute of fading channels referred in this paper indicates that the channel power gains are random, but remain invariant for a sufficiently large period of time \cite{tse2005fundamentals}. This aligns with the practical scenarios of the IoT with stationary MTDs, which are subject to slow fading and supported by proper power supply\footnote{There are many application scenarios of the IoT with stationary MTDs in practice, which have been summarized in the 3GPP TSG-RAN WG2 Meeting Report (Huawei), including smart metering, telemedicine, environment monitoring, and home automation \cite{huawei}. In this report, it is found that for a series of IoT services, MTDs are stationary and are capable of utilizing the configuration parameters obtained by an initial setup procedure and updating them by a small-size data packet. This stationary attribute is also assumed in a series of published papers associated with the IoT and MTC \cite{7354795,8055645,6883918,7468557}.}. Owing to the slow fading assumption, the signaling overheads rendered by performing subcarrier assignment and codebook feedforward to the receiver for detection purposes become negligible \cite{5752793,8241721}.

\section{Outage Performance Analysis}\label{opa}
\subsection{Definition of Average Outage Probability}
To analyze the reliability of enhanced OFDM-SNM, we define the subcarrier-wise  conditional outage probability conditioned on SAP $k$ for the $n$th subcarrier in the first place. This probability refers to the occurrence of the event that the received SNR $\gamma(k,n)$ of an arbitrary active subcarrier $n\in\mathcal{N}(k)$ is smaller than a preset outage threshold $\xi$, which is mathematically given by
\begin{equation}\label{tengs98721}
\begin{split}
\Phi(k,n)&=\mathbb{P}\left\lbrace \gamma(k,n)<\xi\right\rbrace=\mathbb{P}\left\lbrace |h(n)|^2<\frac{T(k)N_0\xi}{P_t}\right\rbrace\\
&=F_g\left(\frac{T(k)N_0\xi}{P_t}\right),
\end{split}
\end{equation}
where $\mathbb{P}\left\lbrace\cdot\right\rbrace$ denotes the probability of the random event enclosed.

For modern multi-carrier communication systems, e.g., OFDM, it is common that the information borne over multiple active subcarriers has certain correlations for error detection and/or correction, etc., which impose a much more stringent requirement on multi-carrier signal detection than that of single-carrier systems. Specifically, it is required that all subcarriers in use must be well received with higher SNRs than the preset outage threshold $\xi$, for both conventional OFDM systems \cite{6157252,8344837,8110602} and OFDM-IM systems \cite{8361430,8476574,8612925}. As a result, we have the following definition of the conditional outage probability considering all active subcarriers for SAP $k$:
\begin{equation}\label{chaikebaier23231}
\Phi(k)=\mathbb{P}\left\lbrace\underset{n\in\mathcal{N}(k)}{\bigcup}\left\lbrace\gamma(k,n)<\xi\right\rbrace\right\rbrace.
\end{equation}
To capture the fact that different numbers of active subcarriers $T(k)$ will affect the outage performance, we remove the condition on SAP $k$ by averaging over all SAPs and define the average outage probability to be
\begin{equation}\label{djsahj2}
\overline{\Phi}=\underset{k}{\mathbb{E}}\left\lbrace\Phi(k)\right\rbrace,
\end{equation}
which is used for measuring the outage probability of the proposed system utilizing enhanced OFDM-SNM.

\subsection{Derivation of Average Outage Probability}
First of all, we can reduce (\ref{chaikebaier23231}) by fundamental probability theory for the finite union relation and obtain
\begin{equation}\label{closj22222daska}
\begin{split}
\Phi(k)&=1-\prod_{n\in\mathcal{N}(k)}\left(1-\Phi_{\{l_n\}}(k,n)\right),
\end{split}
\end{equation}
where $\Phi_{\{l_n\}}(k,n)$ is the subcarrier-wise conditional outage probability when the $n$th subcarrier is ranked as the $l_n$th smallest in terms of instantaneous channel power gain $|h(n)|^2$. To derive the average outage probability, we should first focus on two scenarios when the enhanced OFDM-SNM is in use, depending on whether all subcarriers are activated. This is because $\Phi_{\{l_n\}}(k,n)$ is related to subcarrier assignment by enhanced OFDM-SNM. We discuss both scenarios in the following paragraphs.

\subsubsection{$T(k)<N$}
According to the system model described in Section \ref{sm}, when $T(k)<N$, subcarrier assignment will be employed to activate $T(k)$ subcarriers so as to maximize the sum of their instantaneous channel power gain. By (\ref{dasjd22226a}), it can be easily found that the subcarrier assignment is equivalent to activating the $T(k)$ subcarriers with the $(N-T(k)+1)$th to the $N$th smallest instantaneous channel power gains $|h(n)|^2$. Because the outage event is associated with the worst active subcarrier with the $(N-T(k)+1)$th smallest instantaneous channel power gain, we can resort to order statistics and simplify (\ref{closj22222daska}) to be \cite{david2004order}
\begin{equation}\label{meiquankai}
\begin{split}
\Phi(k)\vert_{T(k)<N}&=\sum_{n=N-T(k)+1}^{N}\binom{N}{n}\left(F_g\left(\frac{T(k)N_0\xi}{P_t}\right)\right)^n\\
&~~~~~~~~~~~~~~~~~~\times\left(1-F_g\left(\frac{T(k)N_0\xi}{P_t}\right)\right)^{N-n},
\end{split}
\end{equation}
where $\binom{\cdot}{\cdot}$ denotes the binomial coefficient.

\subsubsection{$T(k)=N$}
According to the system model described in Section \ref{sm}, when $T(k)=N$, we do not need to perform subcarrier assignment. Therefore, all active subcarriers can be regarded as homogeneous, which simplifies (\ref{closj22222daska}) to be
\begin{equation}\label{quankaile}
\begin{split}
\Phi(k)\vert_{T(k)=N}&=1-\prod_{n=1}^{N}\left(1-\Phi(k,n)\right)\\
&=1-\left(1-F_g\left(\frac{NN_0\xi}{P_t}\right)\right)^N.
\end{split}
\end{equation}

Eventually, by the mapping relation between incoming bit sequences and SAPs specified in Section \ref{sm} as well as the equiprobable property of incoming bits, the average outage probability defined in (\ref{djsahj2}) can be determined by
\begin{equation}\label{tajssss21931023}
\overline{\Phi}=\sum_{\zeta=1}^{N}\Upsilon(\zeta)\Phi(k)\vert_{T(k)=\zeta}=\frac{1}{N}\sum_{\zeta=1}^{N}\Phi(k)\vert_{T(k)=\zeta},
\end{equation}
where $\Upsilon(\zeta)=\mathbb{P}\left\lbrace T(k)=\zeta\right\rbrace=1/N$ denotes the probability that the number of active subcarriers is $\zeta$.

\subsection{Power Series Expansion on Average Outage Probability at High SNR}
In order to illustrate the relation among average outage probability and crucial system parameters, we perform power series expansion on average outage probability for large SNR ($P_t/N_0\rightarrow\infty$) and aim at obtaining the asymptotic expression. Similarly, as what we derived the average outage probability, we analyze the scenarios depending on $T(k)$ as follows.

\subsubsection{$T(k)<N$}
By (\ref{dkasjdk2pdfcdf}),  we can reduce (\ref{meiquankai}) by the binomial theorem in (\ref{diyichangtxiaoyundenage}) at the top of the next page, where $_2F_1(a,b,c;z)$ is the Gauss hypergeometric function \cite{m2014asymptotic}.

\begin{figure*}[!t]
\begin{equation}\label{diyichangtxiaoyundenage}
\begin{split}
\Phi(k)\vert_{T(k)<N}\sim\tilde{\Phi}(k)\vert_{T(k)<N}=\left(\frac{T(k)N_0\xi}{P_t\mu}\right)^{N-T(k)+1}\binom{N}{N-T(k)+1} {_2F_1}\left(1,1-T(k),N-T(k)+2;-\frac{T(k)N_0\xi}{P_t\mu}\right)
\end{split}
\end{equation}
\hrule
\end{figure*}

\subsubsection{$T(k)=N$}
We can similarly perform the same methodology as for the case of $T(k)<N$ and derive the asymptotic expression of (\ref{quankaile}) to be
\begin{equation}\label{lingwaiasympt}
\begin{split}
\Phi(k)\vert_{T(k)=N}\sim\tilde{\Phi}(k)\vert_{T(k)=N}=\frac{N^2N_0\xi}{P_t\mu}.
\end{split}
\end{equation}

Thereafter, substituting (\ref{diyichangtxiaoyundenage}) and (\ref{lingwaiasympt}) into (\ref{tajssss21931023}) yields the asymptotic expression of the average outage probability at high SNR, from which it is clear that no diversity gain can be harvested by subcarrier assignment, but a coding gain is provided in comparison to the original OFDM-SNM published in \cite{8362748}. This can be easily shown by $d_o=-\underset{P_t/N_0\rightarrow\infty}{\lim}\left\lbrace\frac{\log\left(\Phi(k)\vert_{T(k)=N}\right)}{\log(P_t/N_0)}\right\rbrace=1$. This unity-diversity-order system can be explained as follows. According to the fundamentals of wireless communications \cite{proakis2008digital}, diversity techniques can be viewed as the supply of multiple replicas of the same information-bearing signal by different orthogonal paths in a variety of signal domains. It is also observed that the outage performance of multi-carrier systems is dominated by the worst active subcarrier with the lowest channel power gain over all legitimate SAPs in the codebook \cite{5456054}. That is, the diversity gain in the frequency domain is produced by the prevention of using `bad' subcarrier(s) in the codebook. However, it is obvious that there still exist SAPs in the optimized codebook after performing subcarrier assignment that activate all subcarriers (c.f. Table \ref{exmtable} for an example where there exist $|\mathcal{X}|=30$ legitimate SAPs, from which 16 SAPs activate all subcarriers). As a consequence, no diversity gain can be harvested by the enhanced OFDM-SNM based on subcarrier assignment.

\section{Error Performance Analysis}\label{epa}
\subsection{Definition of Average Block Error Rate}
Apart from outage performance, error performance is also a key indicator of system reliability and worth investigating for  enhanced OFDM-SNM. However, because the length of entire bit sequence consisting of heading and subsequent bit sequences is variable, bit error rate (BER), a conventional error performance metric, might not be appropriate anymore. That is, an erroneously decoded bit sequence could have a longer or shorter length than the correct one, which results in a difficulty to define the error event in a bit-wise manner. To circumvent confusion and complicated discussion on this issue, we consider the error event in the block level and employ average BLER as the metric to measure error performance for enhanced OFDM-SNM \cite{8614439}.  More specifically, we express the conditional BLER conditioned on instantaneous CSI as
\begin{equation}\label{conditionalerrorpob3662}
P_e\left(\mathbf{x}(k)\vert \mathbf{H}\right)=\mathbb{P}\left\lbrace\hat{\mathbf{x}}(\hat{k})\neq \mathbf{x}(k)\vert \mathbf{H}\right\rbrace.
\end{equation}
Subsequently, we can obtain the unconditional BLER by averaging $P_e\left(\mathbf{x}(k)\vert \mathbf{H}\right)$ over $\mathbf{H}$:
\begin{equation}\label{dsadhaskjh22oohz}
P_e\left(\mathbf{x}(k)\right)=\underset{\mathbf{H}}{\mathbb{E}}\left\lbrace P_e\left(\mathbf{x}(k)\vert \mathbf{H}\right)\right\rbrace,
\end{equation}
which characterizes the error performance for the OFDM block $\mathbf{x}(k)$. To cover all legitimate OFDM blocks $\mathbf{x}(k)\in\mathcal{X}$ and investigate the error performance on a comprehensive basis, it is straightforward to average $P_e\left(\mathbf{x}(k)\right)$ over $\mathbf{x}(k)$ and finally have the average BLER:
\begin{equation}\label{dsakdjk221finsdef}
\bar{P}_e=\underset{\mathbf{x}(k)\in\mathcal{X}}{\mathbb{E}}\left\lbrace P_e\left(\mathbf{x}(k)\right)\right\rbrace,
\end{equation}
which we employ in this paper to evaluate the error performance of enhanced OFDM-SNM.

\subsection{Approximation of Average Block Error Rate}

To derive the average BLER, we first need to pay attention to and derive its basic element, the conditional BLER conditioned on instantaneous CSI, i.e., $P_e\left(\mathbf{x}(k)\vert \mathbf{H}\right)$. To do so, we can employ the classic methodology involving pairwise error probability (PEP) analysis to help with the derivation and approximate (\ref{conditionalerrorpob3662}) to be \cite{7544555,8006223}
\begin{equation}\label{dsakdjkasjdk223}
P_e\left(\mathbf{x}(k)\vert \mathbf{H}\right)\approx\sum_{\hat{\mathbf{x}}(\hat{k})\neq\mathbf{x}(k)}P_e\left(\mathbf{x}(k)\rightarrow \hat{\mathbf{x}}(\hat{k})\vert\mathbf{H}\right),
\end{equation}
where $P_e\left(\mathbf{x}(k)\rightarrow \hat{\mathbf{x}}(\hat{k})\vert\mathbf{H}\right)$ represents the conditional PEP conditioned on instantaneous CSI $\mathbf{H}$ quantifying the probability that the originally transmitted OFDM block $\mathbf{x}(k)$ is erroneously estimated to $\hat{\mathbf{x}}(\hat{k})$ at the receiver. With the help of Gaussian tail function (a.k.a. the Q-function) $Q(x)=\frac{1}{\sqrt{2\pi}}\int_{x}^{\infty}\mathrm{exp}\left(-\frac{u^2}{2}\right)\mathrm{d}u$, the conditional PEP $P_e\left(\mathbf{x}(k)\rightarrow \hat{\mathbf{x}}(\hat{k})\vert\mathbf{H}\right)$ can be written as \cite{8640102}
\begin{equation}\label{dsads5a452453232}
\begin{split}
&P_e\left(\mathbf{x}(k)\rightarrow \hat{\mathbf{x}}(\hat{k})\vert\mathbf{H}\right)\\
&=Q\left(\sqrt{\frac{P_t}{N_0}\begin{Vmatrix}\mathbf{H}\left(\frac{\mathbf{x}(k)}{\sqrt{T(k)}}-\frac{\hat{\mathbf{x}}(\hat{k})}{\sqrt{T(\hat{k})}}\right)\end{Vmatrix}^2_F}\right)\\
&=Q\left(\sqrt{\frac{P_t}{N_0}\sum_{n=1}^{N}|h(n)|^2\left\lvert\frac{x(k,n)}{\sqrt{T(k)}}-\frac{x(\hat{k},n)}{\sqrt{T(\hat{k})}}\right\rvert^2}\right)\\
&=Q\left(\sqrt{\frac{P_t}{N_0}\sum_{n=1}^{N}G(n)\Delta(n,k,\hat{k})}\right),
\end{split}
\end{equation}
where we denote $G(n)=|h(n)|^2$ and $\Delta(n,k,\hat{k})=\left\lvert{x(k,n)}/{\sqrt{T(k)}}-{x(\hat{k},n)}/{\sqrt{T(\hat{k})}}\right\rvert^2$ for simplicity. Because by the original definition of Gaussian tail function, the argument is the lower limit of the interval of an integral, it is thereby difficult to perform further analysis. To solve this difficulty, we adopt an exponential approximation of $Q(x)\approx \sum_{i=1}^2 \rho_i\mathrm{exp}(-\eta_ix^2)$, where $\{\rho_1,\rho_2\}=\left\lbrace\frac{1}{12},\frac{1}{4}\right\rbrace$ and $\{\eta_1,\eta_2\}=\left\lbrace\frac{1}{2},\frac{2}{3}\right\rbrace$ \cite{1210748}, and approximate (\ref{dsads5a452453232}) in an alternative form as
\begin{equation}\label{craigseqqpro}
\begin{split}
&P_e\left(\mathbf{x}(k)\rightarrow \hat{\mathbf{x}}(\hat{k})\vert\mathbf{H}\right)\\
&\approx\sum_{i=1}^{2}\rho_i\mathrm{exp}\left(-\frac{\eta_iP_t}{N_0}\sum_{n=1}^{N}G(n)\Delta(n,k,\hat{k})\right)\\
&=\sum_{i=1}^{2}\rho_i\prod_{n=1}^{N}\mathrm{exp}\left(-\frac{\eta_iP_tG(n)\Delta(n,k,\hat{k})}{N_0}\right).
\end{split}
\end{equation}
In order to remove the condition on $\mathbf{H}$ and obtain $P_e\left(\mathbf{x}(k)\right)$, by taking advantage of the additivity of expectation operation, we approximate the relation infra from (\ref{dsadhaskjh22oohz}) and (\ref{dsakdjkasjdk223}):
\begin{equation}
\begin{split}
&P_e\left(\mathbf{x}(k)\right)=\underset{\mathbf{H}}{\mathbb{E}}\left\lbrace P_e\left(\mathbf{x}(k)\vert \mathbf{H}\right)\right\rbrace\\
&\approx\underset{\mathbf{H}}{\mathbb{E}}\left\lbrace\sum_{\hat{\mathbf{x}}(\hat{k})\neq\mathbf{x}(k)}P_e\left(\mathbf{x}(k)\rightarrow \hat{\mathbf{x}}(\hat{k})\vert\mathbf{H}\right)\right\rbrace\\
&=\sum_{\hat{\mathbf{x}}(\hat{k})\neq\mathbf{x}(k)}\underset{\mathbf{H}}{\mathbb{E}}\left\lbrace P_e\left(\mathbf{x}(k)\rightarrow \hat{\mathbf{x}}(\hat{k})\vert\mathbf{H}\right)\right\rbrace.
\end{split}
\end{equation}
Now, let us focus on the derivation of $\underset{\mathbf{H}}{\mathbb{E}}\left\lbrace P_e\left(\mathbf{x}(k)\rightarrow \hat{\mathbf{x}}(\hat{k})\vert\mathbf{H}\right)\right\rbrace$. Again, as different signaling procedures will be applied depending on different $T(k)$, we should discuss the cases for $T(k)<N$ and $T(k)=N$, respectively:
\subsubsection{$T(k)<N$}
When $T(k)<N$, subcarrier assignment is applied and order statistics should be involved to derive the average BLER. By (\ref{dkasjdk2pdfcdf}), the PDF of the $\upsilon$th order statistic of the instantaneous channel power gain among $N$ subcarriers can be written as \cite{david2004order}
\begin{equation}\label{orderstastiticpdf23231}
\phi_{\langle\upsilon\rangle}(\nu)=\frac{N!(F_g(\nu))^{\upsilon-1}(1-F_g(\nu))^{N-\upsilon}f_g(\nu)}{(\upsilon-1)!(N-\upsilon)!}.
\end{equation}
To facilitate the calculation involving subcarrier assignment and subchannel ordering, we rearrange $\mathbf{x}(k)$ by the \textit{orders} of subcarriers rather than the \textit{indices} \cite{8519769}. Then, we can obtain the \textit{permuted} OFDM block\footnote{One should note that the concept of permuted OFDM block introduced in analytical derivations is simply for facilitating the error performance analysis and expression of results, but will not amend the actual transmission procedure of enhanced OFDM-SNM.} $\mathbf{z}(k)=[x(k,\lambda_1),x(k,\lambda_2),\dots,x(k,\lambda_N)]^T\in\mathbb{C}^{N\times 1}$, so that such a relation is validated: $G(\lambda_1)<G(\lambda_2)<\dots<G(\lambda_N)$. By involving the concept of the permuted OFDM block, for $T(k)<N$ we can express (\ref{craigseqqpro}) in an alternative form:
\begin{equation}\label{sasasu23123125552}
\begin{split}
&P_e\left(\mathbf{z}(k)\rightarrow \hat{\mathbf{z}}(\hat{k})\vert\mathbf{H}\right)\\
&=\sum_{i=1}^{2}\rho_i\prod_{\upsilon=1}^{N}\mathrm{exp}\left(-\frac{\eta_iP_tG(\lambda_\upsilon)\Delta(\lambda_\upsilon,k,\hat{k})}{N_0}\right).
\end{split}
\end{equation}
With the help of (\ref{orderstastiticpdf23231}) and (\ref{sasasu23123125552}), we derive $\underset{\mathbf{H}}{\mathbb{E}}\left\lbrace P_e\left(\mathbf{x}(k)\rightarrow \hat{\mathbf{x}}(\hat{k})\vert\mathbf{H}\right)\right\rbrace$ for $T(k)<N$ in (\ref{orderchanggongshi}) at the top of the next page, where $(\mathrm{a})$ is derived by the independence among subcarriers by the system model assumed in this paper; $\Gamma(x)=\int_{0}^{\infty}u^{x-1}\mathrm{exp}(-u)\mathrm{d}u$ is the Gamma function.
\begin{figure*}[!t]
\begin{equation}\label{orderchanggongshi}
\begin{split}
&\underset{\mathbf{H}}{\mathbb{E}}\left\lbrace P_e\left(\mathbf{x}(k)\rightarrow \hat{\mathbf{x}}(\hat{k})\vert\mathbf{H}\right)\right\rbrace\vert_{T(k)<N}=\underset{\mathbf{H}}{\mathbb{E}}\left\lbrace P_e\left(\mathbf{z}(k)\rightarrow \hat{\mathbf{z}}(\hat{k})\vert\mathbf{H}\right)\right\rbrace\\
&=\underbrace{\int_{0}^{\infty}\int_{0}^{\infty}\dots\int_{0}^{\infty}}_{N-\mathrm{fold}}\left(\sum_{i=1}^{2}\rho_i\prod_{\upsilon=1}^{N}\mathrm{exp}\left(-\frac{\eta_iP_tG(\lambda_\upsilon)\Delta(\lambda_\upsilon,k,\hat{k})}{N_0}\right)\right)\left(\prod_{\upsilon=1}^{N}\phi_{\langle \upsilon\rangle}(G(\lambda_\upsilon))\right)\mathrm{d}G(\lambda_1)\mathrm{d}G(\lambda_2)\dots\mathrm{d}G(\lambda_N)\\
&\overset{(\mathrm{a})}{=}\sum_{i=1}^{2}\left(\rho_i\prod_{\upsilon=1}^{N}\int_{0}^{\infty}\mathrm{exp}\left(-\frac{\eta_iP_tG(\lambda_\upsilon)\Delta(\lambda_\upsilon,k,\hat{k})}{N_0}\right)\phi_{\langle \upsilon\rangle}(G(\lambda_\upsilon))\mathrm{d}G(\lambda_\upsilon)\right)\\
&=\sum_{i=1}^{2}\left(\rho_i\prod_{\upsilon=1}^{N}\frac{N!\Gamma\left(N-\upsilon+1+\frac{\eta_iP_t\mu\Delta(\lambda_\upsilon,k,\hat{k})}{N_0}\right)}{(N-\upsilon)!\Gamma\left(N+1+\frac{\eta_iP_t\mu\Delta(\lambda_\upsilon,k,\hat{k})}{N_0}\right)}\right)
\end{split}
\end{equation}
\hrule
\end{figure*}

\subsubsection{$T(k)=N$}
When $T(k)=N$, since all subcarriers are required to be active, there is no need to perform subcarrier assignment and order statistics is thereby not applied. By (\ref{dkasjdk2pdfcdf}) and (\ref{craigseqqpro}), we perform the derivation in (\ref{tknexactevags22312}) at the top of the next page, where $(\mathrm{a})$ is again derived by the independence among subcarriers.

\begin{figure*}[!t]
\begin{equation}\label{tknexactevags22312}
\begin{split}
&\underset{\mathbf{H}}{\mathbb{E}}\left\lbrace P_e\left(\mathbf{x}(k)\rightarrow \hat{\mathbf{x}}(\hat{k})\vert\mathbf{H}\right)\right\rbrace\vert_{T(k)=N}\\
&=\underbrace{\int_{0}^{\infty}\int_{0}^{\infty}\dots\int_{0}^{\infty}}_{N-\mathrm{fold}}\left(\sum_{i=1}^{2}\rho_i\prod_{n=1}^{N}\mathrm{exp}\left(-\frac{\eta_iP_tG(n)\Delta(n,k,\hat{k})}{N_0}\right)\right)\left(\prod_{n=1}^{N}f_g(G(n))\right)\mathrm{d}G(1)\mathrm{d}G(2)\dots\mathrm{d}G(N)\\
&\overset{(\mathrm{a})}{=}\sum_{i=1}^{2}\left(\rho_i\prod_{n=1}^{N}\int_{0}^{\infty}\mathrm{exp}\left(-\frac{\eta_iP_tG(n)\Delta(n,k,\hat{k})}{N_0}\right)f_g(G(n))\mathrm{d}G(n)\right)=\sum_{i=1}^{2}\left[\rho_i\prod_{n=1}^{N}\left(1+\frac{\eta_iP_t\mu\Delta(n,k,\hat{k})}{N_0}\right)^{-1}\right]
\end{split}
\end{equation}
\hrule
\end{figure*}

Finally, according to (\ref{dsakdjk221finsdef}) and the equiprobable property of incoming bits, the average BLER is determined by
\begin{equation}\label{zuihoucuowulv}
\bar{P}_e=\sum_{\mathbf{x}(k)\in\mathcal{X}}\Omega(\mathbf{x}(k))P_e\left(\mathbf{x}(k)\right),
\end{equation}
where $\Omega(\mathbf{x}(k))=1/(NM^{T(k)})$ denotes the probability that the OFDM block $\mathbf{x}(k)$ is in use. Again, it is obvious that only a coding gain is obtainable by enhanced OFDM-SNM with subcarrier assignment, and no diversity gain can be harvested. This result complies with the findings summarized in \cite{1221802} that the average outage probability and average error rate share the identical diversity order but with a shift in SNR.

\section{Numerical Results and Discussions}\label{nr}
\subsection{Verification of Analysis}
\begin{figure}[!t]
\centering
\includegraphics[width=3.5in]{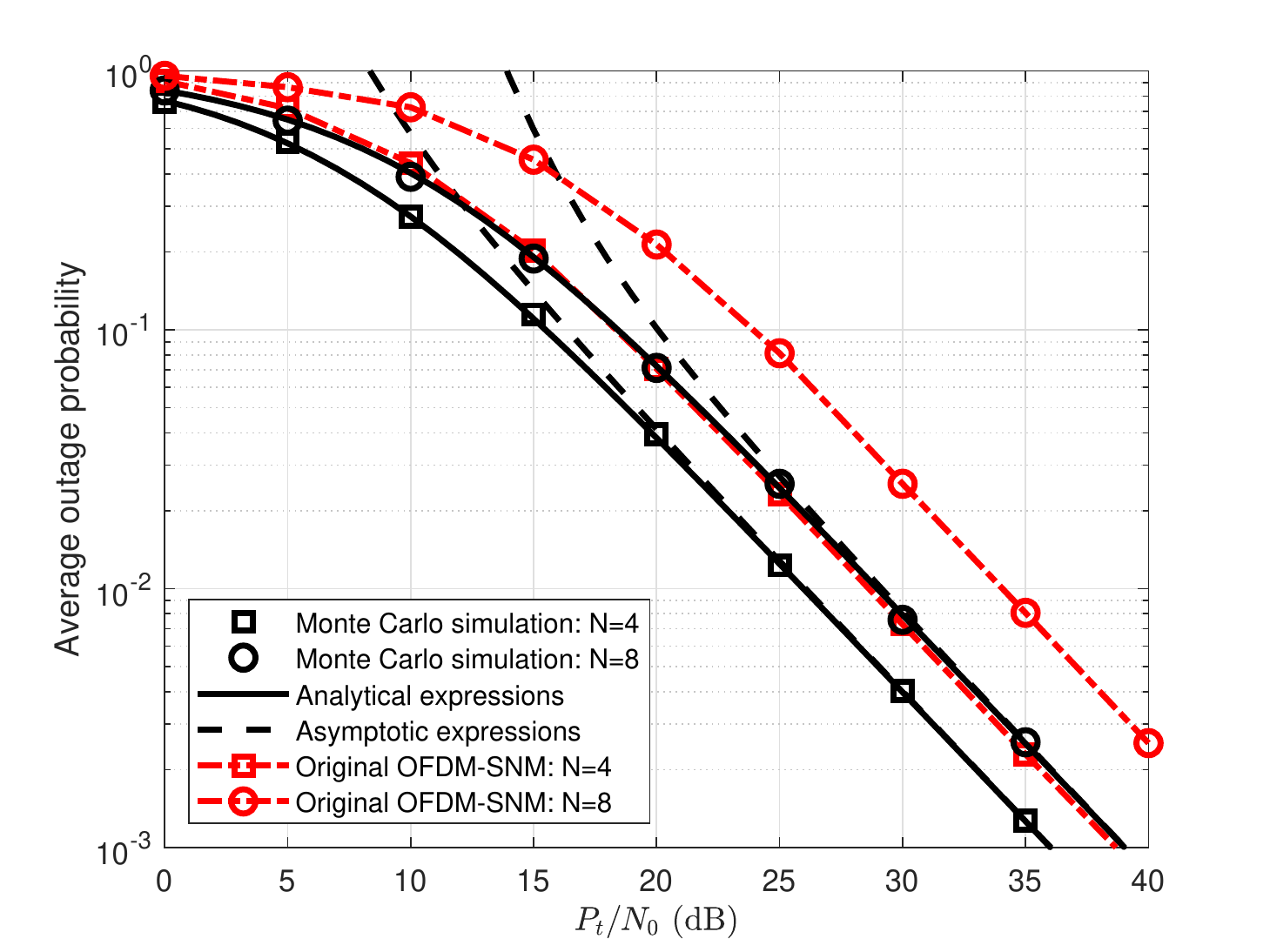}
\caption{Average outage probability vs. ratio of transmit power to noise power $P_t/N_0$.}
\label{outage_verification}
\end{figure}

\begin{figure}[!t]
\centering
\includegraphics[width=3.5in]{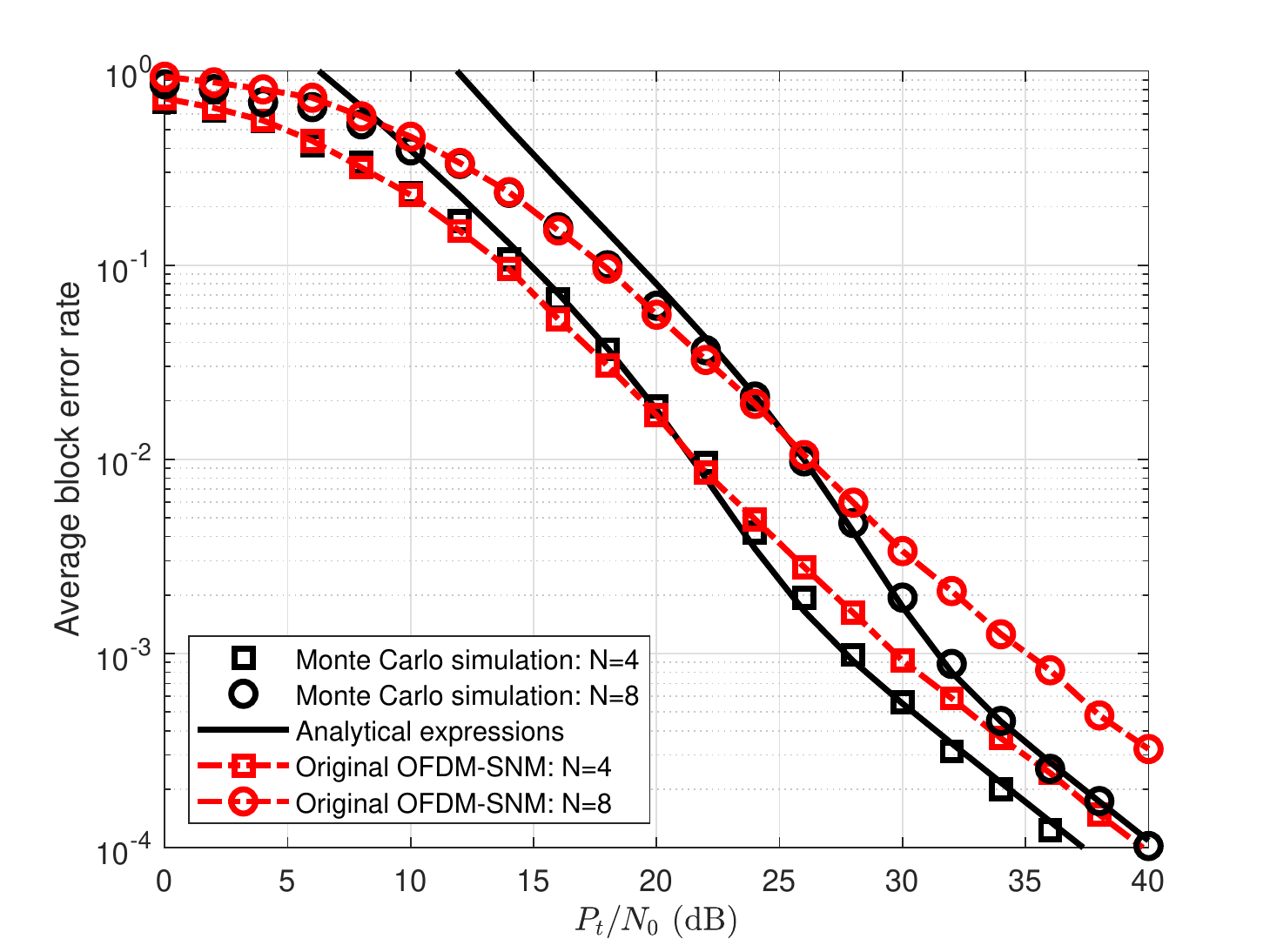}
\caption{Average BLER vs. ratio of transmit power to noise power $P_t/N_0$.}
\label{BLER_verification}
\end{figure}

To verify the outage and error performance analysis presented in Section \ref{opa} and Section \ref{epa}, we carried out a series of simulations by Monte Carlo methods and present the generated numerical results to compare with our analytical results in this section. It should be noted that to maintain the generality, we do not specify the application scenario of these simulations, and normalize most parameters, which gives $\xi=1$ and $\mu=1$. Also, we adopt BPSK as the APM scheme for data constellation symbol carried on each active subcarrier. To illustrate the performance superiority of enhanced OFDM-SNM over original OFDM-SNM without getting access to instantaneous CSI and applying subcarrier assignment, we adopt the original OFDM-SNM published in \cite{8362748} as the performance comparison benchmark in all simulations. The simulation results associated with average outage probability and average BLER are presented in Fig. \ref{outage_verification} and Fig. \ref{BLER_verification}, respectively, with different number of subcarriers $N$. We discuss our discoveries from the simulation results illustrated in both figures as follows.

First of all, from Fig. \ref{outage_verification}, it is obvious that the analytical and asymptotic expressions for average outage probability given in (\ref{tajssss21931023}), (\ref{diyichangtxiaoyundenage}) and (\ref{lingwaiasympt}) have been substantiated. The analytical results well match the numerical results, and the asymptotic results get increasingly close to the numerical results when $P_t/N_0$ becomes large. Besides, the superiority of the enhanced OFDM-SNM over original OFDM-SNM in terms of outage performance can also be verified, as evident constructive coding shifts appear for all cases with different $N$. On the other hand, it aligns with our expectation that there is no diversity gain that can be harvested from the implementation of subcarrier assignment. Besides, by scrutinizing Fig. \ref{outage_verification}, one can also know the impacts of the number of subcarriers $N$ on outage performance. That is, an increasing number of subcarriers $N$ will lead to worse outage performance, simply because of the stringent requirement that all active subcarriers must be well received with higher SNRs than the preset outage threshold $\xi$.

\begin{figure*}[!t]
    \centering
    \begin{subfigure}[t]{0.5\textwidth}
        \centering
        \includegraphics[width=3.5in]{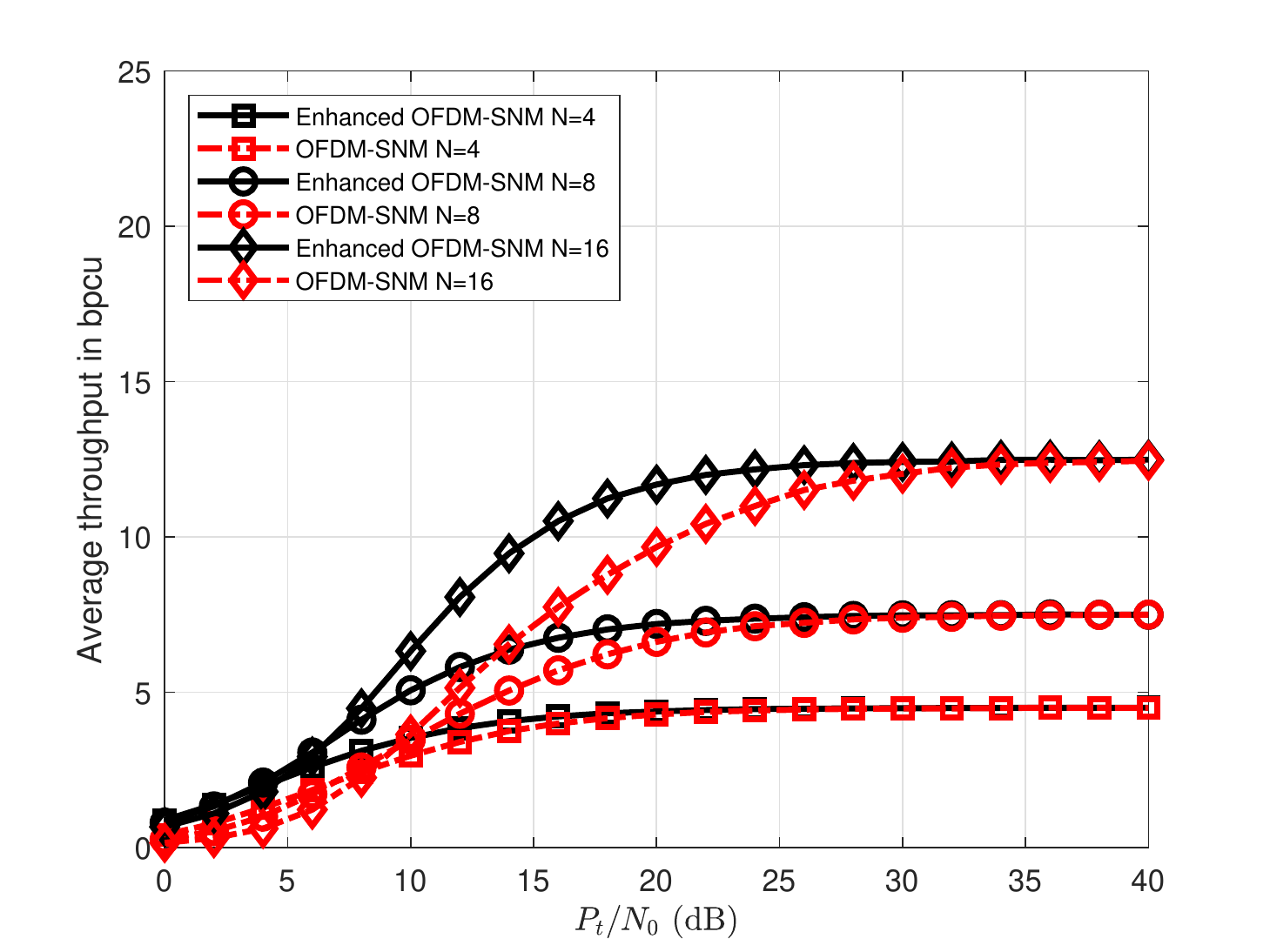}
        \caption{BPSK}
    \end{subfigure}%
~
    \begin{subfigure}[t]{0.5\textwidth}
        \centering
        \includegraphics[width=3.5in]{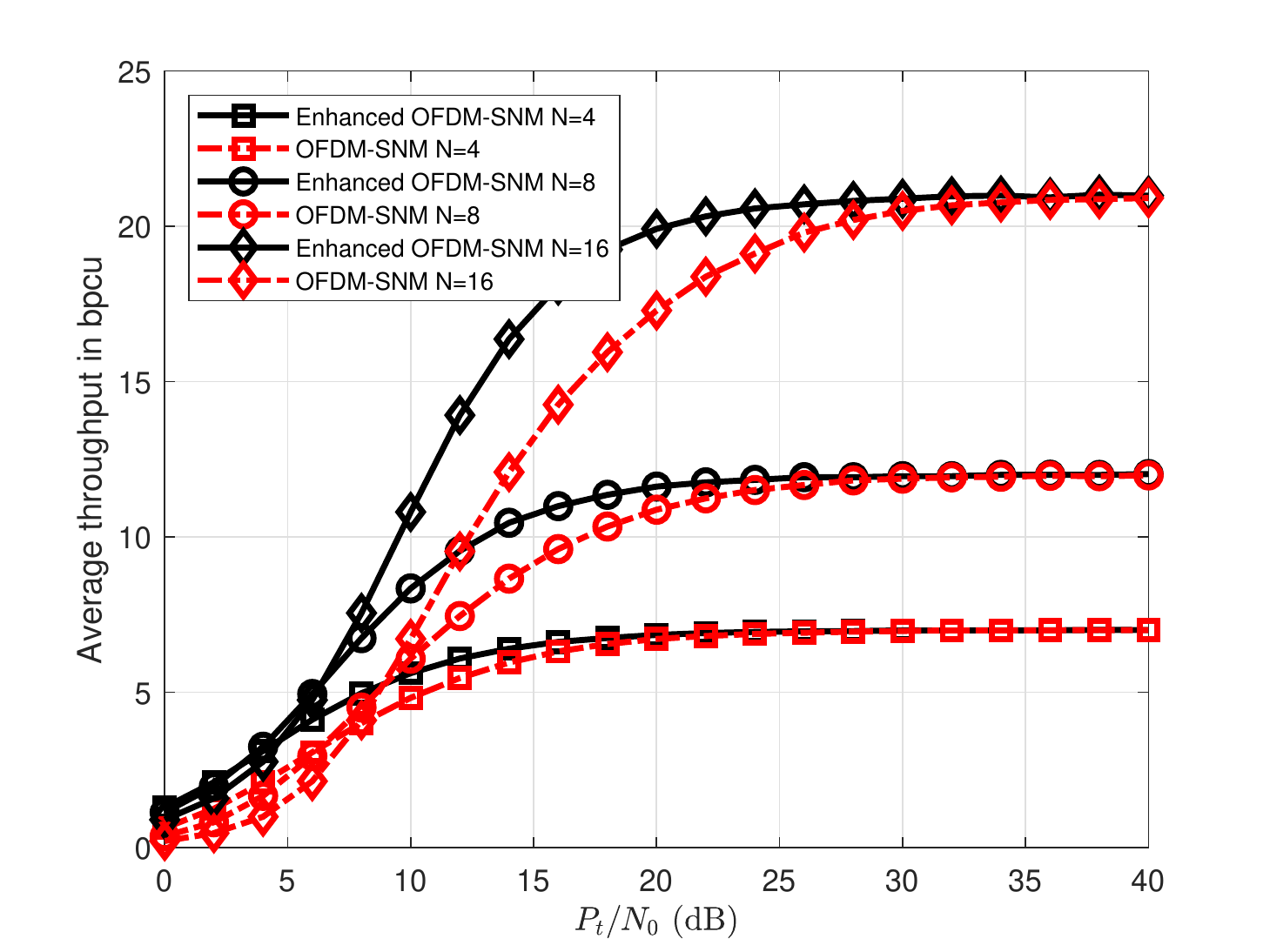}
        \caption{QPSK}
    \end{subfigure}
    \caption{Average throughput vs. ratio of transmit power to noise power $P_t/N_0$.}
    \label{fig_SER_vs_pt}
\end{figure*}

Also, the coding shift between the enhanced OFDM-SNM and its original counterpart will become large when increasing the number of subcarriers $N$. For $N=4$ and $N=8$, the achieved coding gains are ca. 2 dB and 4 dB, respectively. There are two inherent mechanisms that enlarge this coding shift. First, this is because the probability of the occurrence of the special case that all subcarriers are activated (i.e.,  $T(k)=N$) and subcarrier assignment is not applied is $\Upsilon(N)=1/N$, and in this special case, there is no difference in transmission between enhanced OFDM-SNM and original OFDM-SNM. With an increasing number of subcarriers $N$, the occurrence probability of this special case becomes smaller. Second, with a larger number of subcarriers $N$, when $T(k)<N$, more subcarriers are possible to be assigned, which is more likely to find a proper subset of subcarriers according to (\ref{dasjd22226a}), and thereby leads to a more reliable system.

Second, we can also verify the analysis of error performance given in (\ref{zuihoucuowulv}) by observing Fig. \ref{BLER_verification}, as the derived approximate results approach numerical results when the ratio of transmit power to noise power $P_t/N_0$ becomes large. The gap between approximate and numerical results at low SNR is because of the joint effects of the PEP-based union bound and the exponential approximation of Q-function (c.f. (\ref{dsakdjkasjdk223}) and (\ref{craigseqqpro})). The impact of $N$ on average BLER follows the same trend as on average outage probability, which aligns with the findings summarized in \cite{1221802} that the average outage probability and average error rate share the identical diversity order but with a shift at high SNR. 

On the other hand, we also find that a slightly counter-intuitive phenomenon that subcarrier assignment will not always bring a constructive effect on the error performance of OFDM-SNM when the ratio of transmit power to noise power $P_t/N_0$ is small. That is, subcarrier assignment would also enhance the erroneous trials for estimation, which cannot be well distinguished from the correct one when $P_t/N_0$ is small. Particular attention should be paid to this phenomenon and sufficient transmit power should be provided in order to maintain the error performance superiority of the enhanced OFDM-SNM over its original counterpart.

\subsection{Average Throughput}
Besides, to be comprehensive, we also inspect the transmission efficiency of the proposed OFDM-SNM by numerically investigating its average throughput in Fig. \ref{fig_SER_vs_pt} for BPSK and quadrature PSK (QPSK). The same simulation configurations are adopted as for the outage and error performance verification and the original OFDM-SNM is again taken as the comparison benchmark. It can be shown that the proposed enhanced OFDM-SNM outperforms the original OFDM-SNM, and both will converge to the same and invariant average throughput at high SNR. Increasing either the number of subcarriers $N$ or APM order $M$ will result in a larger average throughput. It should be noted that at large SNR, the average throughput approaching the average transmission rate determined in (\ref{sdkasjkdj2}) will not be affected by whether coding techniques and other performance enhancement mechanisms are used and is only dependent on the mapping relation between incoming bit streams and transmission patterns as well as data constellation symbols.

\subsection{Transmit Diversity Scheme}
As we mentioned in the previous sections, there is no diversity gain harvested by the enhanced OFDM-SNM, and only a coding gain can be achieved. However, for applications demanding high reliability, it is also possible to tailor the proposed scheme by inserting a subcarrier halving procedure before the SNM mapper to attain a diversity gain. Specifically, such a subcarrier halving procedure selects $N/2$ subcarriers out of $N$ subcarriers with larger channel power gains and then the proposed OFDM-SNM scheme is performed over these $N/2$ selected subcarriers\footnote{Here, we assume that the number of subcarriers $N$ is a power of two for simplicity, which is a common case in modern multi-carrier systems.}. Note that, the halving procedure involves a sub-channel \textit{ordering} process. As a consequence of such an ordering process, we can harvest a frequency diversity gain. The mechanism of the diversity gain attained in this way is the same as the diversity mechanism by adopting an adaptive subcarrier selection in classic OFDM systems \cite{4022475}. On the other hand, this subcarrier halving procedure will inevitably reduce the average throughput in the high SNR region as a cost of achieving the diversity gain.

\begin{figure*}
    \begin{subfigure}[t]{0.3\textwidth}
        \includegraphics[width=2.5in]{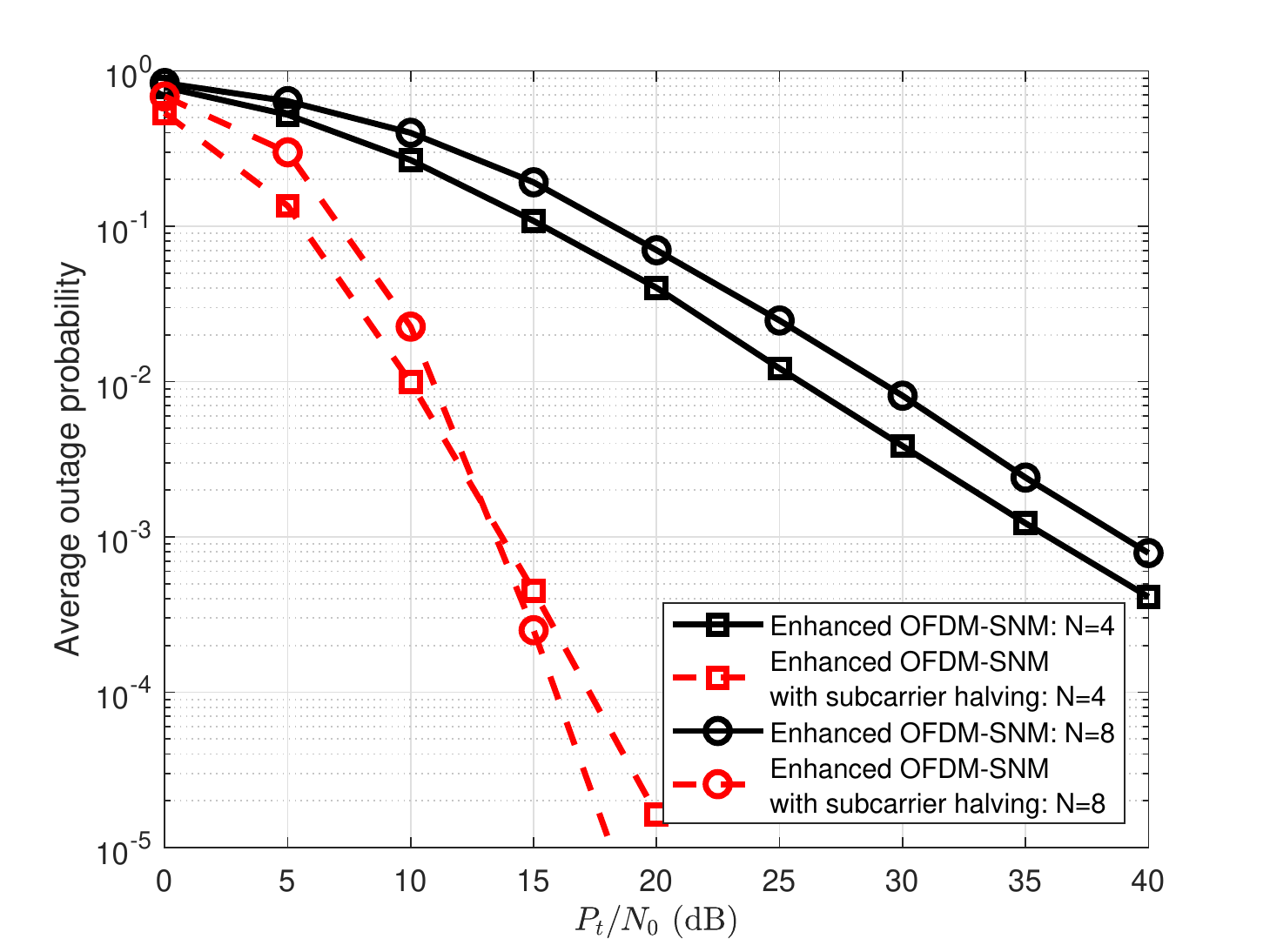}
        \caption{Outage performance}
        \label{compare_outage}
    \end{subfigure}
    ~~ 
    \begin{subfigure}[t]{0.3\textwidth}
        \includegraphics[width=2.5in]{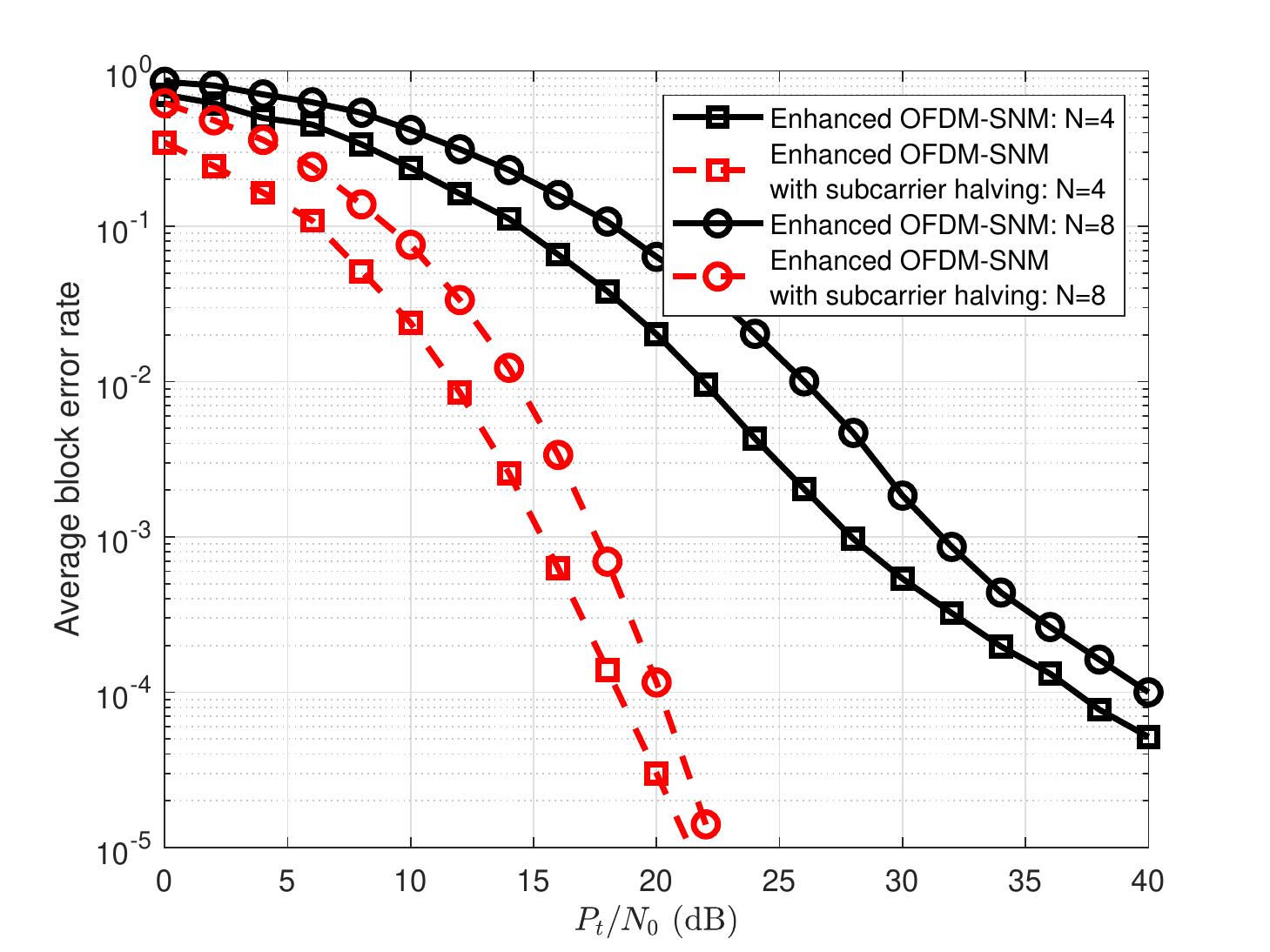}
        \caption{Error performance}
        \label{compare_BLER}
    \end{subfigure}
    ~~ 
    \begin{subfigure}[t]{0.3\textwidth}
        \includegraphics[width=2.5in]{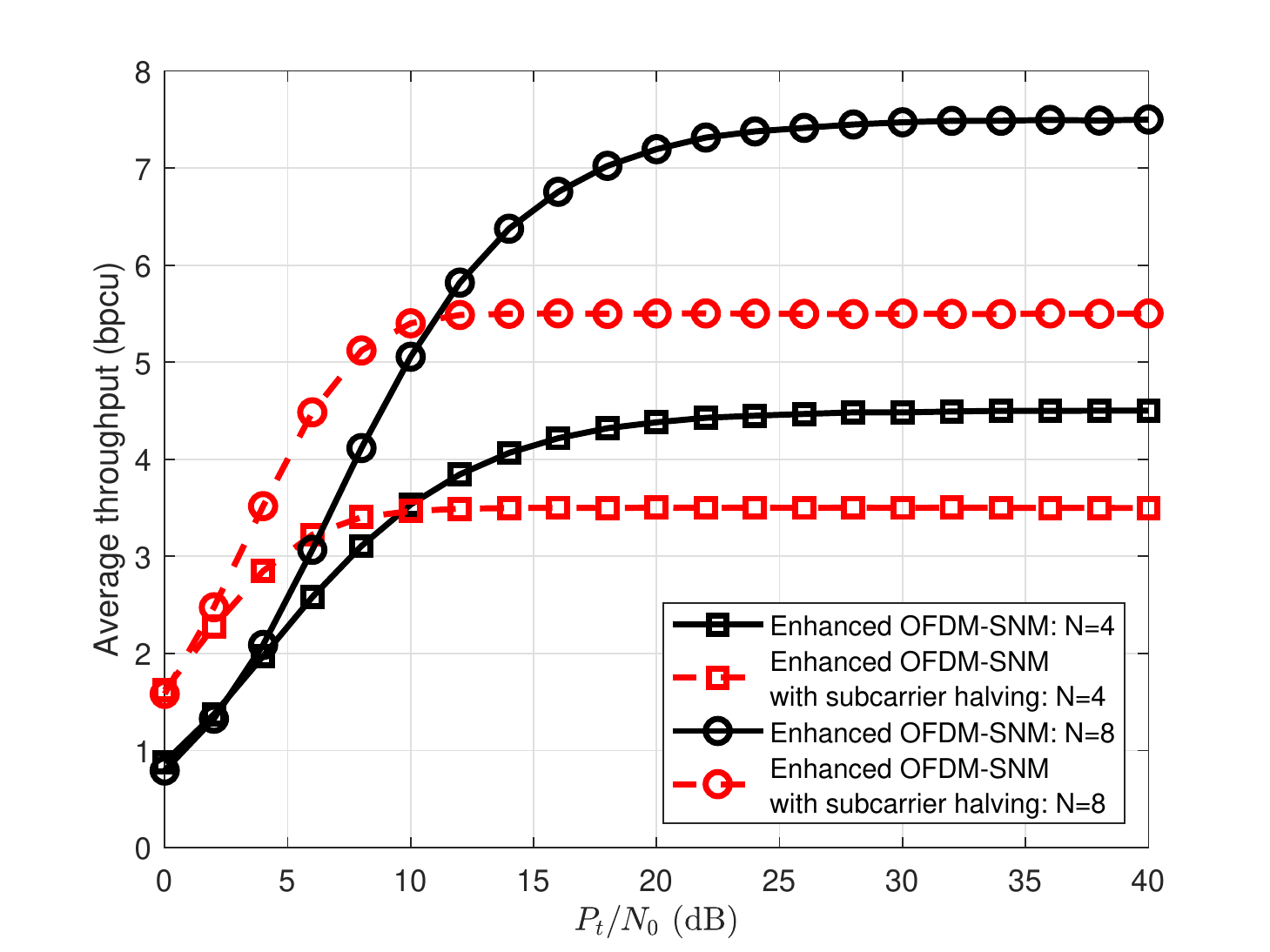}
        \caption{Throughput}
        \label{compare_capacity}
    \end{subfigure}
    \caption{Performance comparisons between enhanced OFDM-SNM systems with and without subcarrier halving when BPSK is in use ($M=2$).}\label{comparison}
\end{figure*}

We illustrate numerical results corresponding to the halved cases with $N=4$ and $N=8$ as well as the benchmarks without the subcarrier halving procedure in Fig. \ref{comparison}. Observing this figure, it is verified that by involving the subcarrier halving procedure, the enhanced OFDM-SNM system is capable of achieving a diversity order of $1+N/2$, since the red curves corresponding to the outage and error performance of enhanced OFDM-SNM systems with the subcarrier halving procedure decay much faster with an increasing $P_t/N_0$ due to higher diversity orders. However, red curves corresponding to the throughput of enhanced OFDM-SNM systems with the subcarrier halving procedure are lower than their black counterparts at high SNR.

\subsection{Multi-User Scenario}
As multi-user scenarios are more realistic and of high importance for practical wireless communication systems, we examine the feasibility of our proposed enhanced OFDM-SNM scheme in multi-user scenarios in this subsection. Here, we consider a biased multi-user architecture \cite{7577711}, where there are one primary user applying the proposed enhanced OFDM-SNM scheme and $L$ secondary users that are able to perfectly sense the idle subcarriers not being used by the primary user and use them for their own transmissions. Obviously, such a transmission architecture will cause interference to the receiver of the primary user, while the network throughput considering multiple users would be increased. We assume that the channel power gains corresponding to the transmission and interference channels of all secondary users are independently and exponentially distributed with different average channel power gains $\varrho$ and $\theta$, respectively. We further suppose that all secondary users are homogeneous and transmit by the same power $P_s$. Two transmission protocols for secondary users are considered in the simulations:
\begin{itemize}
\item \textbf{Unregulated transmission protocol}: A idle subcarrier that is not activated and used by the primary user via the enhanced OFDM-SNM scheme will be used by one of the $L$ secondary users that generates the lowest interference.
\item \textbf{Regulated transmission protocol}: A idle subcarrier that is not activated and used by the primary user via the enhanced OFDM-SNM scheme will be used by one of the $L$ secondary users that generates the lowest interference \textit{if and only if} the lowest generated interference is lower than a preset threshold $\phi$.
\end{itemize}

\begin{figure}[!t]
\centering
\includegraphics[width=3.5in]{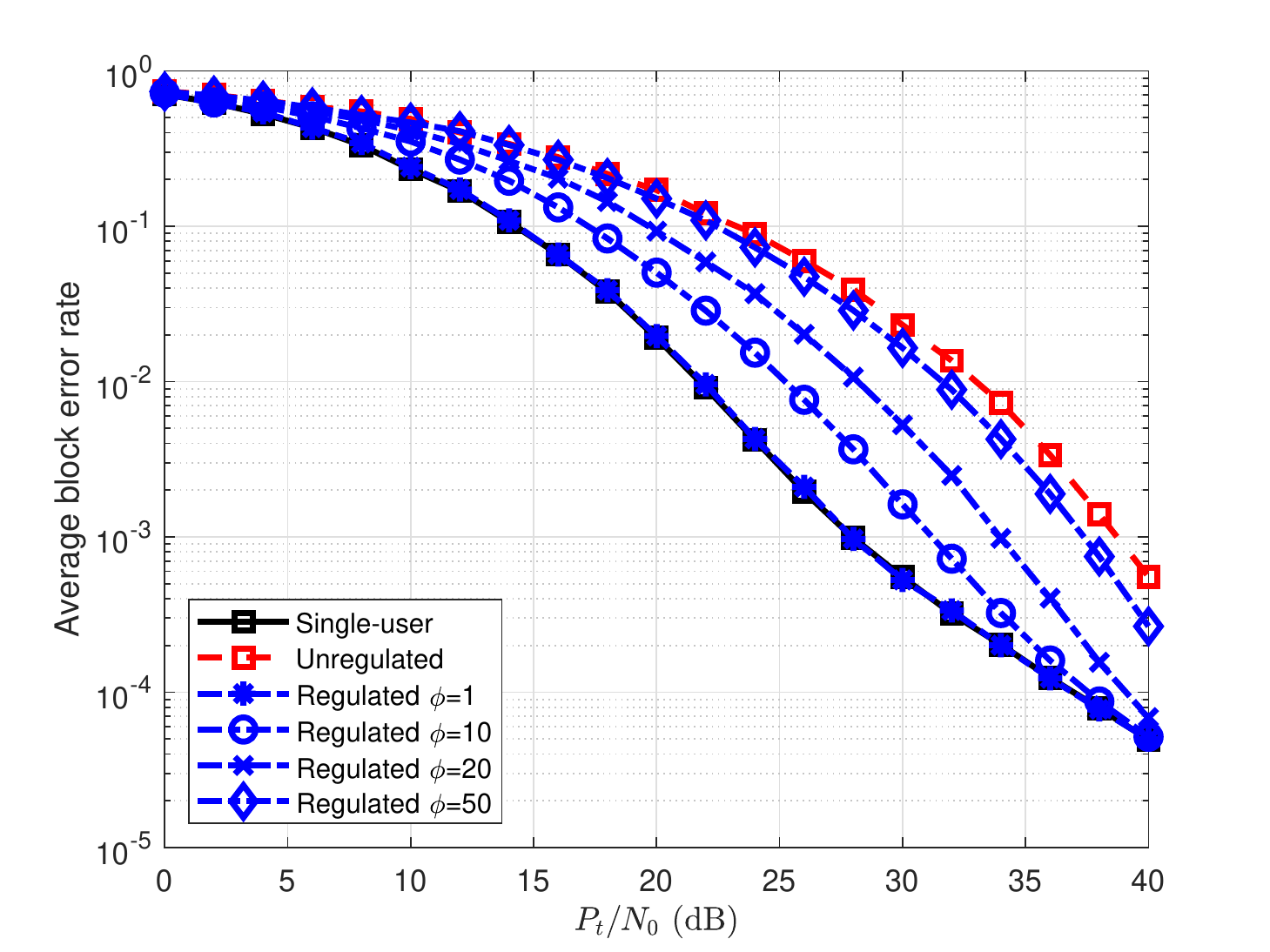}
\caption{Average BLER in single-user scenario and various multi-user scenarios utilizing different transmission protocols.}
\label{multiusersimu}
\end{figure}

For simplicity, we let $\varrho=1$, $\theta=0.2$, $L=1$ (referring to the integrated node model introduced in \cite{8344837}), and $P_s/N_0=20$ dB. The numerical results regarding the error performance of the enhanced OFDM-SNM scheme in the single-user scenario and various multi-user scenarios are demonstrated in Fig. \ref{multiusersimu}. From this figure, it is verified that the transmissions of secondary users will have a negative impact on the error performance of the primary user applying the enhanced OFDM-SNM scheme, which leads to a higher average BLER. Specifically, when $\phi\rightarrow 0$, all secondary transmissions are terminated, and the regulated transmission scenario is reduced to the single-user case. When $\phi\rightarrow\infty$, the performance of the multi-user systems abiding the regulated transmission protocol converges to the performance of the multi-user systems employing the unregulated transmission protocol. Furthermore, it is evident that when increasing $P_t/N_0$, the average BLERs corresponding to all cases will converge. This is because when $P_t\gg P_s$, the effect of the secondary transmissions on the primary transmission becomes negligible.

\section{Conclusion}\label{c}
To enhance the system reliability of OFDM-SNM, a fresh `sibling' scheme of OFDM-IM, we proposed an enhanced OFDM-SNM scheme in this paper. The enhanced OFDM-SNM exploits the flexibility of placing subcarriers and performs subcarrier assignment to harvest a coding gain. Specifically, we stipulated a subcarrier assignment scheme relying on instantaneous CSI, which dynamically maps a sequence of bits to an optimized SAP consisting of subcarriers with higher channel power gains. We analyzed the outage and error performance of the proposed system utilizing enhanced OFDM-SNM. The average outage probability and BLER have been derived and approximated in closed-form expressions. Also, the asymptotic expression for average outage probability has been determined, so as to reflect the diversity order of the proposed system. All aforementioned analytical results were corroborated by numerical results generated by Monte Carlo simulations. Numerical results were provided to verify the performance superiority of the enhanced OFDM-SNM over the original OFDM-SNM without implementing subcarrier assignment. Because the machine-type nodes are normally stationary and the channel variation among nodes is less volatile, the additional signaling overhead caused by dynamical optimization on mapping relation between bit sequences and SAPs by subcarrier assignment can be mitigated to a reasonable level, which makes enhanced OFDM-SNM a promising candidate for implementing in the IoT with stationary MTDs.

On the other hand, there exist several questions awaiting solution before implementing enhanced OFDM-SNM in practice, which could be regarded as future research directions. First, the optimization of the number of subcarriers considering reliability, average throughput, and detection complexity is worth investigating. In particular, one can halve the number of subcarriers to achieve a diversity gain at the cost of reduced average throughput. Second, cognitive radio (CR) protocol would be useful to incorporate the enhanced OFDM-SNM into a multi-user framework, which suits more realistic scenarios and is thereby worth further studying. Also, as the transmit power is assumed to be uniformly distributed over all active subcarriers in this paper, well-designed power allocation schemes would be considered as another constructive mechanism to improve the system performance.

\appendix
\section*{Comparison of Data Transmission Rates}\label{cdtr}
Before comparing, we first present the transmission rate of OFDM-IM and plain OFDM infra \cite{6587554}:
\begin{equation}
p_{\mathrm{IM}}=\left\lfloor \log_2\left(\binom{N}{T}\right) \right\rfloor+T\log_2(M)
\end{equation}
and
\begin{equation}\label{ofdmrate}
p_{\mathrm{OFDM}}=N\log_2(M)
\end{equation}
where $1\leq T<N$ is a fixed number of active subcarriers predefined by OFDM-IM.

\subsection{Comparison of Transmission Rates between OFDM-SNM and OFDM-IM}
\begin{figure}[!t]
\centering
\includegraphics[width=3.5in]{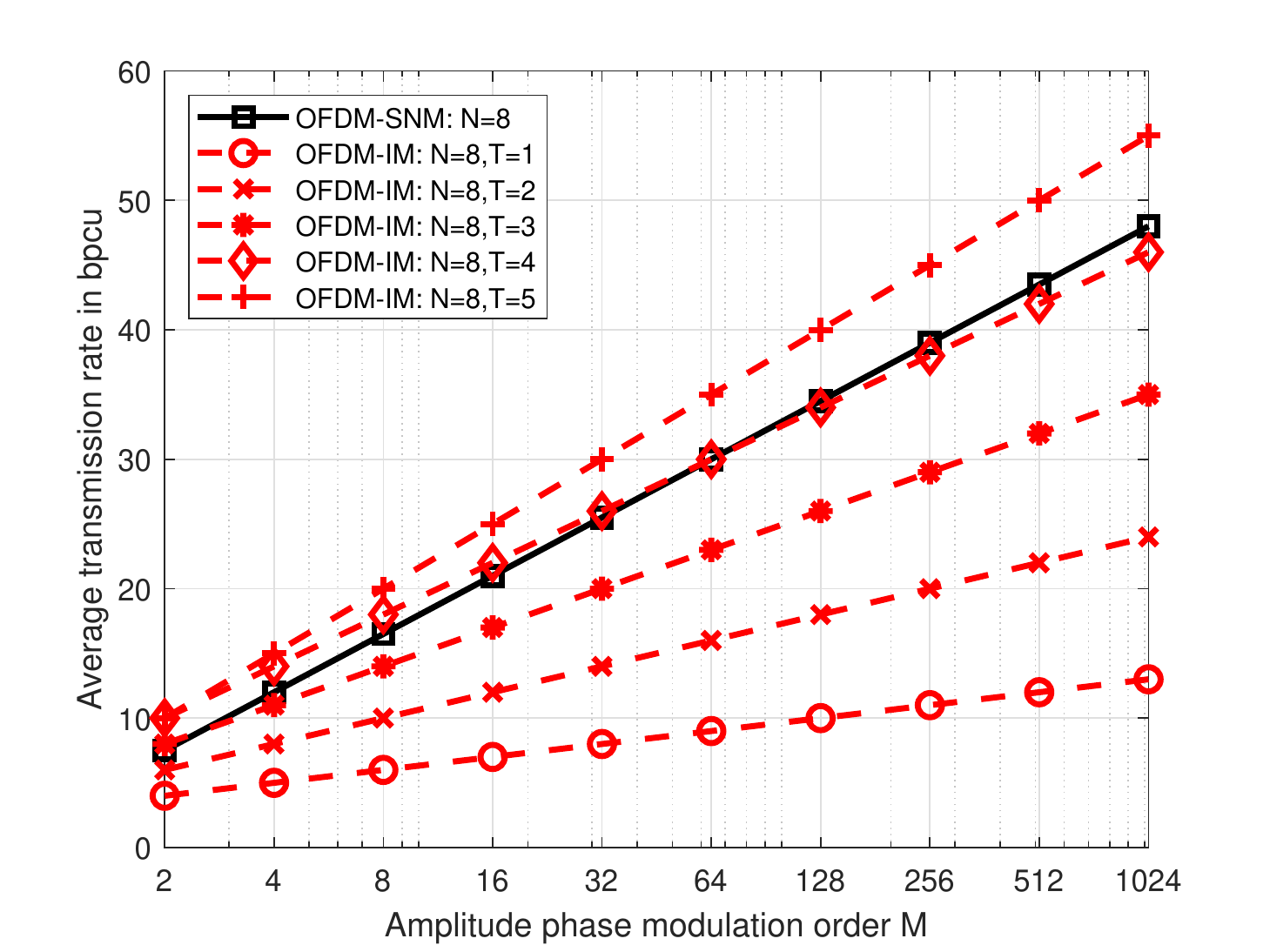}
\caption{Average transmission rate vs. APM order $M$ for OFDM-SNM and OFDM-IM.}
\label{rate_comparison_SNM_IM}
\end{figure}

To provide insightful details of the impacts of $M$, $N$, and $T$ on the transmission rate superiority, we restrict our discussion to the cases when $N$ is a power of two, and thereby the average transmission rate of OFDM-SNM adopts the form given in (\ref{sdkasjkdj10000}) without involving the floor function. Then, we employ an upper bound on $p_{\mathrm{IM}}$ as
\begin{equation}
p_{\mathrm{IM}}\leq \log_2\left(\binom{N}{T}\right)+T\log_2(M).
\end{equation}
Assuming $\bar{p}\geq p_{\mathrm{OFDM}}$, this inequality can be released to
\begin{equation}\label{sdashdj2ineqs}
\log_2(N)+\frac{N+1}{2}\log_2(M)\geq \log_2\left(\binom{N}{T}\right)+T\log_2(M).
\end{equation}
Simplifying (\ref{sdashdj2ineqs}) yields the relation
\begin{equation}\label{dsakdjksa2565dsa6}
    \log_M\left({N}/{\binom{N}{T}}\right) +\frac{N+1}{2}-T\geq 0.
\end{equation}
Because $N/\binom{N}{T}\leq1$, $\log_M\left({N}/{\binom{N}{T}}\right)$ must be a non-positive term, the necessary condition of (\ref{dsakdjksa2565dsa6}) is thereby
\begin{equation}\label{sdasdk2454d}
T\leq \frac{N+1}{2}.
\end{equation}
Let $f(M,N,T)=\log_M\left({N}/{\binom{N}{T}}\right) +\frac{N+1}{2}-T$. Because $N/\binom{N}{K}\leq 1$, $f(M,N,T)$ is a monotone increasing function of $M$ and the range of $M$ achieving (\ref{sdashdj2ineqs}) can be determined as
\begin{equation}
M\geq\left(N/\binom{N}{T}\right)^{\frac{1}{T-\frac{N+1}{2}}}
\end{equation}
conditioned on the satisfaction of (\ref{sdasdk2454d}). Considering $M\geq 2$, we can list the sets of $M$ corresponding to different combinations of $N$ and $T$ in Table \ref{snmimrateM}. For verification purposes, we plot the average transmission rate for both OFDM-SNM and OFDM-IM in Fig. \ref{rate_comparison_SNM_IM}, in which the presented results align with our expectation.

\begin{table}[!t]
\renewcommand{\arraystretch}{1.3}
\caption{Sets of $M$ achieving $\bar{p}\geq p_{\mathrm{IM}}$ corresponding to different combinations of $N$ and $T$.}
\label{snmimrateM}
\centering
\begin{tabular}{C{1cm}|C{1.5cm}|C{1.5cm}|C{1.5cm}}
\hline
\diagbox{$T$}{$N$} & 2 & 4 & 8\\
\hline\hline
1 & $M\geq 2$ & $M\geq 2$ & $M\geq 2$ \\
2 & N/A & $M\geq 4$ & $M\geq 2$\\
3 & N/A & $\varnothing$ & $M\geq 4$\\
4 & N/A & N/A & $M\geq 128$\\
5 & N/A & N/A & $\varnothing$\\
6 & N/A & N/A & $\varnothing$\\
7 & N/A & N/A & $\varnothing$\\
\hline
\end{tabular}
\end{table}

\subsection{Comparison of Transmission Rates between OFDM-SNM and plain OFDM}
\begin{figure}[!t]
\centering
\includegraphics[width=3.5in]{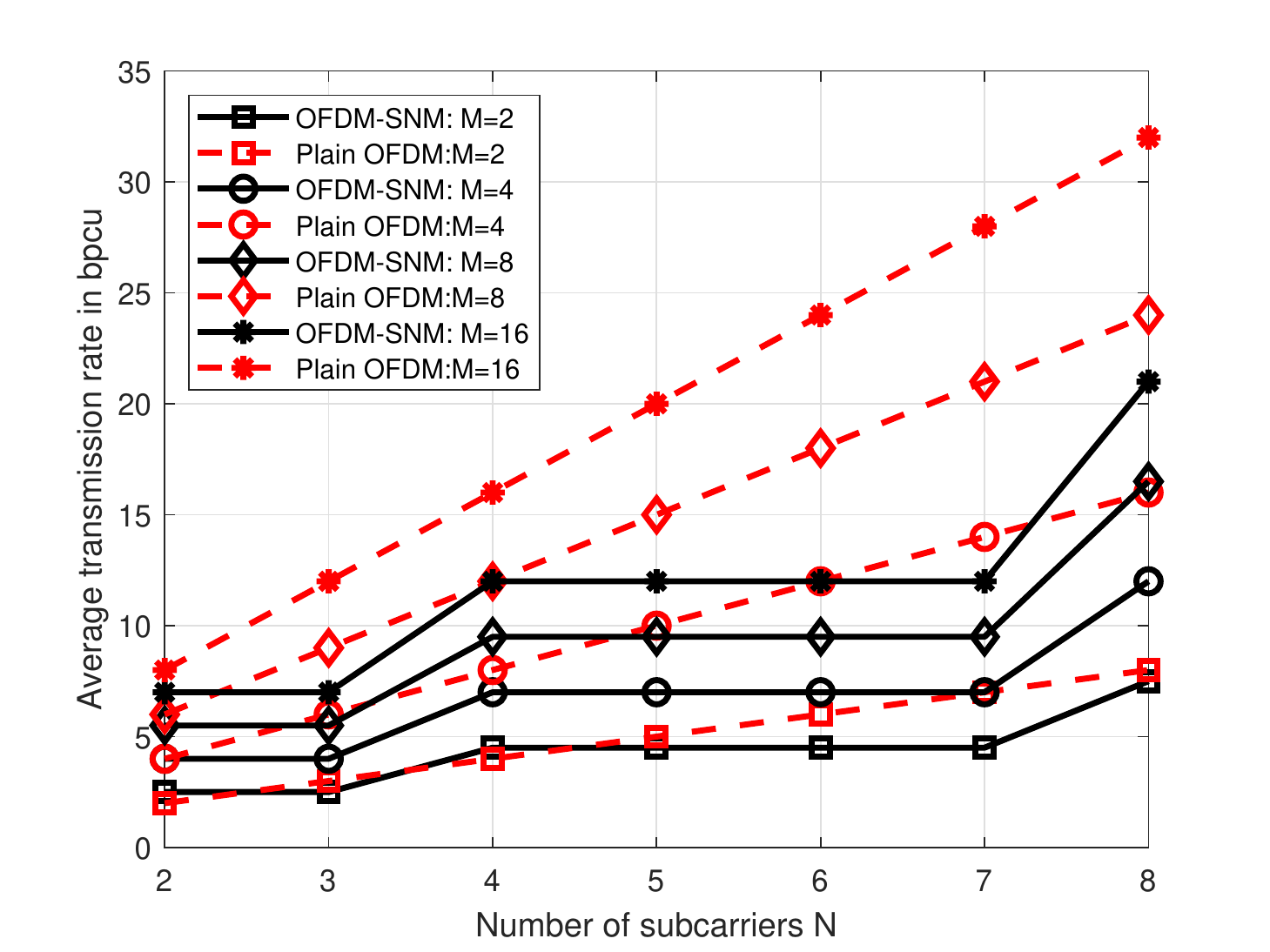}
\caption{Average transmission rate vs. number of subcarriers $N$ for OFDM-SNM and plain OFDM.}
\label{rate_comparison_SNM_Plain_OFDM}
\end{figure}

Again, to provide insightful details of the impacts of $M$ and $N$ on the transmission rate superiority, we restrict our discussion to the cases when $N$ is a power of two, and thereby the average transmission rate of OFDM-SNM adopts the form given in (\ref{sdkasjkdj10000}) without involving the floor function. Subsequently, assuming $\bar{p}\geq p_{\mathrm{OFDM}}$, we can refer to (\ref{sdkasjkdj10000}) and (\ref{ofdmrate}) to deduce the following relation:
\begin{equation}\label{sdkasjk255896s}
\log_2(N)+\frac{N+1}{2}\log_2(M)\geq N\log_2(M)
\end{equation}
which can be further simplified to
\begin{equation}
    \frac{1}{2} + \log_M(N)  - \frac{N}{2} \geq 0.
\end{equation}
Now, let $g(M,N)=\frac{1}{2} + \log_M(N)  - \frac{N}{2}$ and inspect the monotonicity of $g(M,N)$ with respect to $M$. Because $N\geq 2$, it is clear that for a given $N$, $g(M,N)$ is a monotone decreasing function of $M$. In other words, with an increasing $M$, it is less likely that OFDM-SNM has a higher rate than plain OFDM. In particular, $M$ should satisfy the condition 
\begin{equation}
M\leq N^{\frac{2}{N-1}}
\end{equation}
in order to achieve (\ref{sdkasjk255896s}). Therefore, we can list all possible combinations of $N$ and $M$ in Table \ref{NMsnmofdm}. In short, there exist only three combinations $(M,N)=(2,2)$, $(M,N)=(2,4)$, and $(M,N)=(4,2)$ satisfying $\bar{p}\geq p_{\mathrm{OFDM}}$. To verify the above analysis and visually illustrate the rate comparison between OFDM-SNM and plain OFDM, we plot $\bar{p}$ and $p_{\mathrm{OFDM}}$ in Fig. \ref{rate_comparison_SNM_Plain_OFDM}.

\begin{table}[!t]
\renewcommand{\arraystretch}{1.3}
\caption{Possible combinations of $N$ and $M$, by which $\bar{p}\geq p_{\mathrm{OFDM}}$.}
\label{NMsnmofdm}
\centering
\begin{tabular}{C{1cm}|C{1cm}|C{1.5cm}}
\hline
$N$ & $N^{\frac{2}{N-1}}$ & Set of $M$\\
\hline\hline
2 & 4 & $\{2,4\}$\\
4 & $\approx 2.520$ & $\{2\}$\\
8 & $\approx 1.811$ & $\varnothing$\\
\hline
\end{tabular}
\end{table}

\section*{Acknowledgment}
We thank the editor and the anonymous reviewers for their constructive comments, which have helped us improve the quality of the paper. We also appreciate the discussion with Dr. Jehad M. Hamamreh and Mr. Ahmad M. Jaradat with Istanbul Medipol University via emails.

\bibliographystyle{IEEEtran}
\bibliography{bib}

\end{document}